\documentclass[twocolumn,english,showpacs,nofootinbib,superscriptaddress]{revtex4-1}
\pdfoutput=1
\usepackage[T1]{fontenc}
\usepackage[latin9]{inputenc}
\usepackage{geometry}
\usepackage{xcolor}
\usepackage{babel}
\usepackage{units}
\usepackage{amsmath}
\usepackage{amssymb}
\usepackage{mathtools} %Eran added this.
\usepackage{braket} %Eran added this.
\usepackage{stmaryrd}
\usepackage{graphicx}
\usepackage{wasysym}
\usepackage[bookmarks=false]{hyperref}
\usepackage{soul}

\usepackage{tabularx} %Added by Eran for \textwidth table.

\makeatletter
\IfFileExists{lmodern.sty}{\usepackage{lmodern}}{}
\usepackage{babel}
\usepackage{orcidlink}

\usepackage{hyperref}
\def\equationautorefname~#1\null{Equation (#1)\null}

\makeatother

\begin{document}
\title{Optimal Strategies for Optical Quantum Memories Using Long-Lived Noble-Gas Spins}
\author{Or Katz\orcidlink{0000-0001-7634-1993}}
\thanks{These authors contributed equally to this work.}
\address{School of Applied and Engineering Physics, Cornell University, Ithaca, NY 14853.}
\email[Corresponding authors: ]{or.katz@cornell.edu; eran.reches@mpq.mpg.de}

\author{Eran Reches\orcidlink{0000-0001-8358-0095}}
\thanks{These authors contributed equally to this work.}
\affiliation{Department of Physics of Complex Systems, Weizmann Institute of Science,
Rehovot 76100, Israel}
\affiliation{Max-Planck-Institut f{\"u}r Quantenoptik, 85748 Garching, Germany}
\affiliation{Fakult{\"a}t f{\"u}r Physik, Ludwig-Maximilians-Universit{\"a}t M{\"u}nchen, 80799 M{\"u}nchen, Germany}
\author{Roy Shaham}
\affiliation{Department of Physics of Complex Systems, Weizmann Institute of Science,
Rehovot 76100, Israel}

\affiliation{Harvard-MIT Center for Ultracold Atoms, Cambridge, Massachusetts 02138, USA}
\affiliation{Department of Physics and Department of Chemistry and Chemical Biology, Harvard University, Cambridge, Massachusetts 02138, USA}

\author{Eilon Poem}
\affiliation{Department of Physics of Complex Systems, Weizmann Institute of Science,
Rehovot 76100, Israel}

\author{Alexey V. Gorshkov}
\affiliation{Joint Quantum Institute and Joint Center for Quantum Information and
Computer Science, NIST/University of Maryland, College Park, Maryland
20742, USA}

\author{Ofer Firstenberg\orcidlink{0000-0001-8905-9954}}
\affiliation{Department of Physics of Complex Systems, Weizmann Institute of Science,
Rehovot 76100, Israel}
\begin{abstract}
Nuclear spins of noble gases exhibit exceptionally long coherence times and can potentially serve as a long-lived storage medium for quantum information. We analyze and compare the performance of two mechanisms for mapping the quantum state of light onto the collective spin state of noble gases. The first mechanism utilizes collisional exchange with the electronic spin state of metastable noble-gas atoms, while the second relies on spin-exchange collisions with ground-state alkali-metal atoms.
We describe the operation of an optical quantum memory relying on these two mechanisms using a compact model and study strategies that optimize the memory storage efficiency. Through numerical simulations, we identify optimal sequences for storing optical signals with different signal bandwidths and electronic spin relaxation rates. This work highlights the qualitative difference between the two approaches for using noble gases as long-lived quantum memories at non-cryogenic conditions and outlines the regimes in which they are expected to be efficient.
\end{abstract}
\maketitle

\section{Introduction}

Optical quantum memories enable the storage and retrieval of non-classical photonic signals. High performance memories are vital for various quantum-optics applications, including quantum communication, entanglement distribution, and universal quantum computation \cite{Tittel-review-nature-photonics,oPTICAL-QUANTUM-COMPUTING-SCIENCE,Heshami-2016,Gisin-repeaters,Polzik-RMP-2010}.
The memory storage duration is ultimately limited by the coherence time of the material state utilized by the memory.

Nuclear spins in the cores of noble gases are enclosed by complete electronic shells which isolate them from the environment \cite{Gemmel-60-hours-coherence-time-He-2010,Walker-RMP-2017,Firstenberg-QND,Sinatra-squeezing,Sinatra-squeezing2}. They can maintain their quantum state for many hours and thus serve as a robust storage medium.
However, their transparency in optical frequencies and lack of direct interaction with light complicate their application as an optical memory.
Electronic spins in atoms, however, can act as mediators, efficiently coupling to photons via the dipole interaction and enabling access to nuclear spins through magnetic-like interaction.  Nuclear spins in noble gases can couple to the electronic spins of another optically accessible atomic ensemble through random collisions. The latter ensemble can be either a noble gas in an electronically excited metastable state, where the coupling occurs via metastability-exchange collisions governed by the Coulomb interaction \cite{Sinatra-2005-Metastable}, or an alkali-metal vapor in its electronic ground state, where the Fermi-contact interaction mediates the coupling \cite{SEOP-Happer-RMP,Firstenberg-Weak-collisions}.
Both collisional mechanisms have been proposed as interfaces for utilizing noble gas spins as long-lived optical quantum memories \cite{Sinatra-2005-Metastable,noble-gas-PRA},  primarily focusing on ultra-low bandwidth signals or configurations where the mediator's relaxation is negligible.

The memory performance depends on a set of input control fields, which shape the response of the spins and determine their efficiency to store or retrieve photons.
For optically accessible memories in a standard $\Lambda$ configuration, optimal control analysis reveals an optimal mapping that enables storage of finite bandwidth optical signals via temporal shaping of an optical control field \cite{Gorshkov-PRL,Gorshkov1}, with recent extensions for high bandwidth pulses \cite{kollath2024fast} or single photon generation schemes \cite{vasilev2010single,utsugi2022gaussian}.
In the absence of spin relaxation, this mapping features a universal memory efficiency, which depends exclusively on the degree of optical coupling (\textit{e.g.},~optical depth or cooperativity).
However, in quantum memories based on nuclear spins, the magnetic-like coupling strength between nuclear and electronic spins is fixed, and can be further accompanied by non-negligible relaxation of the electron spins.
Consequently, the standard optimal storage protocols, which, \textit{e.g.},~rely on temporal shaping of the coupling strengths, are no longer applicable.

\begin{figure*}[t]
\begin{centering}
\includegraphics[clip,scale=0.59]{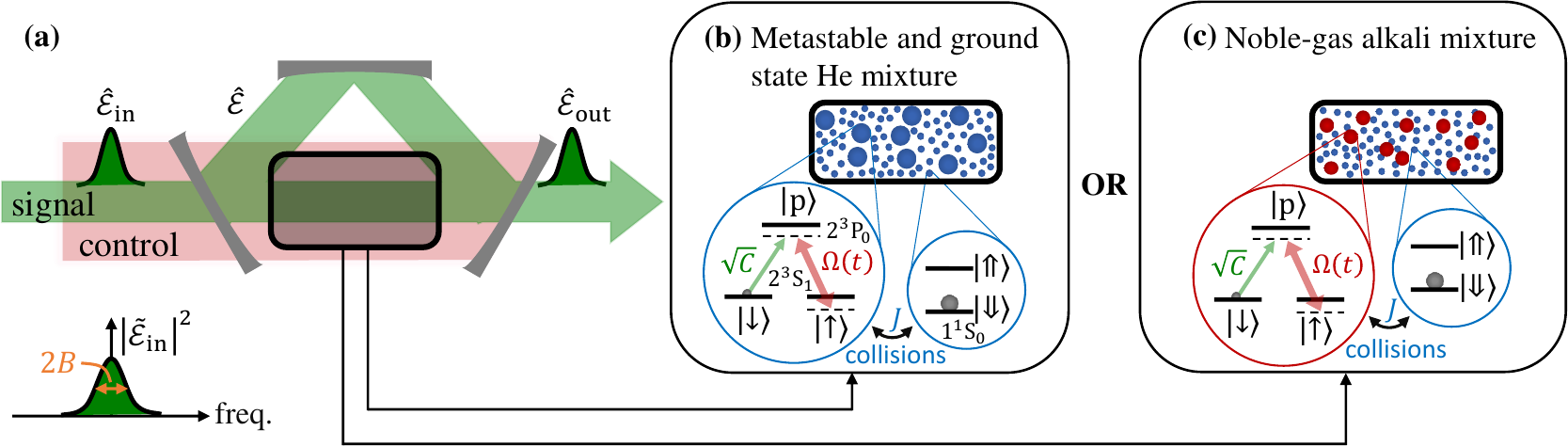}
\par\end{centering}
\centering{}\caption{\textbf{Optical quantum memory using nuclear spins of noble gases.} \textbf{(a)} An input optical
signal (green) is coherently mapped, via electronic spins, onto the state of
noble-gas spins. 
$\hat{\mathcal{E}}$ denotes the annihilation operator of the optical field in the cavity. 
%with bandwidth $2B$. 
$\hat{\mathcal{E}}_{\text{in}}$ ($\hat{\mathcal{E}}_{\text{out}}$) is the annihilation operator for the input (output) optical field. The input pulse has a bandwidth $2B$ where $\tilde{\mathcal{E}}_\text{in}$ is the Fourier transform of the input pulse. Our simple model qualitatively describes platforms where the electronic spins are either \textbf{(b)} of noble gases in a metastable state, \textit{e.g.}~$^3$He, or \textbf{(c)} of alkali-metal atoms in their electronic ground state.
$\Omega(t)$ is the Rabi frequency of the optical control field. $J$ is the fixed, magnetic-like coupling between the nuclear and electronic spins; it originates from metastability exchange collisions in \textbf{(b)}, and from spin-exchange collisions in \textbf{(c)}. \label{fig:setup}}
\end{figure*}

Here, we study the storage of light onto the collective nuclear spin of an ensemble of noble gas atoms. We numerically analyze optimal strategies and compare the two main approaches for such a memory, which are based on either metastability-exchange or spin-exchange collisions with excited-state noble gases or alkali-metal atoms, respectively. For both approaches, we study numerically optimized solutions for variable optical bandwidth of the signal and nonzero relaxation of the mediator, extending previous analyses. We observe that the metastability-exchange approach demonstrates highly efficient storage only for low-bandwidth signals, where the signal is protected via mapping onto a decoherence-free subspace, whereas at higher bandwidths, optimal pulses can achieve moderate efficiencies via non-adiabatic solutions. For the case of spin-exchange collisions, our analysis extends previous analytical protocols and shows that efficient memory operation can be realized for a large range of bandwidths and magnetic-like coupling strengths. Our study delineates the necessary conditions for implementing efficient, hours-long quantum memories under non-cryogenic conditions.

The paper is organized as follows: in Sec.~\ref{Sec:Model}, we present the model studied in this work, consisting of an ensemble of noble-gas spins coupled via collisions to an ensemble of electron spins, which in turn interact with a single mode of an optical cavity. In Sec.~\ref{Sec:Metastable}, we study analytically storage of light based on metastability-exchange collisions with an electronically excited noble gas. In Sec.~\ref{Sec:Spin-Exchange}, we present the main analytical results of Ref.~\cite{noble-gas-PRA} for storage of light based on spin-exchange collisions with alkali-metal vapor. In Sec.~\ref{Sec:numerical}, we extend these analyses and present numerical optimization of the storage sequences for the two different memory configurations. Finally, in Sec.~\ref{Sec:numerical}, we compare the two configurations and in Sec.~\ref{Sec:summary} discuss the results.

\section{Model}\label{Sec:Model}
In this section, we present the model studied in this work: an ensemble of noble-gas spins coupled through collisions to an ensemble of electron spins, which in turn interacts with a single optical cavity mode.

The system we consider, illustrated in Fig.~\ref{fig:setup}, %(a), 
consists of nuclear spins of noble gases and an ensemble of electronic spins. In Fig.~\ref{fig:setup}(b), the electronic spins correspond to metastable noble-gas atoms, while in Fig.~\ref{fig:setup}(c), they represent alkali-metal atoms. The electronic spins interact with an optical signal field $\hat{\mathcal{E}}$ through the electric dipole interaction.
Each electron spin is modeled as a $\Lambda$-system with two spin levels ($\mid\shortdownarrow\rangle$ and $\mid\shortuparrow\rangle$) in the electronic ground state  and one electronically-excited state $|p\rangle$.
To simplify the model, we work with a single-mode cavity described by the annihilation operator $\hat{\mathcal{E}}$ and coupled to the symmetric, collective, optical dipole described by annihilation operator $\hat{\mathcal{P}}\equiv\tfrac{1}{\sqrt{N_\text{a}}}\sum_a\left|\shortdownarrow\right\rangle _a\langle\textnormal{p}|_a$, where $N_\text{a}\gg1$ is the number of electronic spins.
A classical control field with Rabi frequency $\Omega\left(t\right)$ drives the $|\textnormal{p}\rangle$--$\mid\shortuparrow\rangle$ transition, thereby coupling the signal field to the collective electron spin coherence with annihilation operator $\hat{\mathcal{S}}\equiv\tfrac{1}{\sqrt{N_{\text{a}}}}\sum_{a}\left|\shortdownarrow\right\rangle _{a}\langle\shortuparrow\mid_{a}$ \cite{Gorshkov1}.
This two-photon process circumvents the rapid decoherence of the optical dipole $\hat{\mathcal{P}}$ (\textit{e.g.}, due to spontaneous emission, pressure broadening, or inhomogeneous broadening) and, by modulation of $\Omega\left(t\right)$, allows for the storage of
the optical signal on and its retrieval from the spin $\hat{\mathcal{S}}$.
At this point, $\hat{\mathcal{S}}$ may serve as a quantum memory for $N\ll N_{\text{a}}$ photons, whose lifetime is limited by the relaxation rate $\gamma_{\textnormal{s}}$ of the electron spins \cite{Lukin-PRL-2001,Lukin-RMP-2003,Buchler-GEM,Polzik-coherent-state-memory,Polzik-squeezed-states-memory,Walmsley-cavity,Katz-storage-of-light-2018}.

We consider $N_\text{b}$ spin-1/2 nuclei with down and up spin states $\left|\Downarrow\right\rangle $, $\left|\Uparrow\right\rangle $, as shown in Fig.~\ref{fig:setup}(b,c).
These spins, which we use as our quantum memory, weakly interact with their surroundings.
As a result, their collective spin annihilation operator $\hat{\mathcal{K}}\equiv\tfrac{1}{\sqrt{N_{\text{b}}}}\sum_{b}\left|\Downarrow\right\rangle _{b}\left\langle \Uparrow\right|_{b}$
has extremely low decoherence rate $\gamma_{\textrm{k}}\ll\gamma_{\textrm{s}}$ \cite{Firstenberg-Weak-collisions}. 

When the optical input signal enters the cavity, it interacts with the optical dipole and, in the fast-cavity limit, gives rise to the output field \cite{Gorshkov1}
\begin{equation}
\hat{\mathcal{E}}_{\text{out}}=\hat{\mathcal{E}}_{\text{in}}+i\sqrt{2\gamma_{\textnormal{p}}C}\hat{\mathcal{P}}.\label{eq:Output field}
\end{equation}
$C$ is the cooperativity, which characterizes the atom-photon interaction strength and is proportional to the product of the cavity finesse and the optical depth of the atomic medium \cite{Lahad-2017,Gorshkov2}.
$\gamma_{\textnormal{p}}$ denotes the dephasing rate of the atomic optical dipole, and $\gamma_{\textnormal{p}}C$ corresponds to its stimulated emission rate \cite{noble-gas-PRA}. In this work, we focus on the regime of $\gamma_{\textnormal{p}}C\gg B$, where the optical dipole $\hat{\mathcal{P}}$ follows adiabatically the input pulse $\hat{\mathcal{E}}_{\text{in}}$ with bandwidth $2B$ and the optical dipole is weakly excited $\langle\hat{\mathcal{P}}^{\dagger}\hat{\mathcal{P}}\rangle\ll 1$.

The two systems differ in terms of the nature of the electronic spins and the coupling mechanism between the electronic and nuclear spins. However, for spin-polarized ensembles, we can describe the dynamics of both approaches using the Bloch-Heisenberg-Langevin model in a concise manner \cite{HLM1,HLM2}.
While the complete equations, which include quantum noise terms, are stochastic, the memory efficiency and bandwidth are governed by the deterministic part, which is described by\footnote{The stochastic part is mainly responsible for the preservation of
commutation relations through the  introduction of vacuum noise \cite{Polzik-RMP-2010,Firstenberg-Weak-collisions,Gorshkov1}. Specific noise terms for this model are provided in Refs.~\cite{noble-gas-PRA,shaham2020quantum}.}
\begin{align}
\partial_{t}\hat{\mathcal{P}} & =-(\gamma_{\textnormal{p}}(1+C)+i\Delta)\hat{\mathcal{P}}+i\Omega\hat{\mathcal{S}}+i\sqrt{2\gamma_{\textnormal{p}}C}\hat{\mathcal{E}}_{\text{in}},\label{eq:P_coherent_dynamics}\\
\partial_{t}\hat{\mathcal{S}} & =-(\gamma_{\textnormal{s}}+i\delta_{\textnormal{s}})\hat{\mathcal{S}}+i\Omega^{*}\hat{\mathcal{P}}-iJ\hat{\mathcal{K}},\label{eq:S_coherent_dynamics}\\
\partial_{t}\hat{\mathcal{K}} & =-(\gamma_{\textnormal{k}}+i\delta_{\textnormal{k}})\hat{\mathcal{K}}- i\xi J\hat{\mathcal{S}}.\label{eq:K_coherent_dynamics}
\end{align}
Here $\Delta$ denotes the (single-photon) detuning of the atomic optical
transition from the cavity resonance, $\delta_{\textnormal{s}}$ is
the Raman (two-photon) detuning, and $\delta_{\textnormal{k}}$ is
the detuning of the entire three-step process. $J$ denotes the collective magnetic-like coupling rate between the electronic and nuclear spin ensembles.
The parameter $\xi=\mp1$ differentiates between the two experimental configurations we study:
    $\xi=-1$ describes metastability-exchange dynamics for collisions with noble-gas atoms in a metastable state (see Appendix~\ref{appendix:meta_stable_noble_gasses} for derivation based on Ref.~\cite{Sinatra-2005-Metastable}), whereas $\xi=1$ describes the spin-exchange dynamics for collisions with alkali-metal atoms, as derived in Ref.~\cite{Firstenberg-Weak-collisions}. We note that, since Eqs.~(\ref{eq:Output field}-\ref{eq:K_coherent_dynamics}) are linear in the operators, the results presented in the context of storage efficiency are independent of the number of photons in the signal ($N$). Therefore, we set $N = 1$ for the remainder of the paper.

We aim to find and characterize a controllable and reversible process that efficiently transfers the quantum excitations from $\hat{\mathcal{E}}_{\text{in}}$
to the nuclear spins $\hat{\mathcal{K}}$ and then back from $\hat{\mathcal{K}}$ to $\hat{\mathcal{E}}_{\text{out}}$.
In the following, it is important to note that the parameter $\gamma_{\textrm{s}}$ is not negligible, and the coupling $J$ between $\hat{\mathcal{S}}$ and $\hat{\mathcal{K}}$ remains constant, which limits the applicability of the temporal control typically utilized in $\Lambda$-type memories \cite{Gorshkov1}. 

\section{Memories based on Metastability-exchange collisions}\label{Sec:Metastable}
In this section, we analytically study the storage of light based on metastability exchange collisions with an electronically excited noble gas. We identify the decoherence free subspace associated with the metastability exchange coupling that is suitable for the memory operation and estimate the storage efficiency in the adiabatic regime.

The nuclear spin of noble-gas atoms in the electronic ground state can be coupled with the spins of atoms in a metastable electronically excited state through collisions. The ensemble of metastable atoms is typically generated and maintained using pulses of electrical discharge. While the study of various noble-gas atoms has been conducted \cite{lefevre1977metastability,xia2010polarization}, in this work, we focus on an ensemble of $^3$He atoms, which is commonly considered in practical applications \cite{gentile2017optically,batz2011fundamentals,Sinatra-squeezing2}. We refer to the electronic ground-state manifold as $1^{1}\text{S}_{0}$ and the electronic metastable manifold as $2^{3}\text{S}_{1}$. The latter can be controlled through optical transitions and can couple to the former via metastability-exchange collisions, where the exchange of electronic configurations occurs due to strong exchange interaction.

We adopt the assumptions presented in Ref.~\cite{Sinatra-2005-Metastable}, which analyzed storage and retrieval of squeezed light, and consider the fraction $r$ of helium atoms populating the metastable state  to be much smaller than the population in the ground state ($r\ll 1$). We assume that the dominant relaxation mechanism for the helium atoms is the metastability-exchange process, neglecting all other relaxation mechanisms. Under these assumptions, the relaxation rates can be expressed as $\gamma_{\textrm{s}}$ for the metastable population and $\gamma_{\textrm{k}} = r \gamma_{\textrm{s}}$ for the ground-state population. The magnetic-like coupling rate is given by $J = \sqrt{\gamma_{\textrm{k}} \gamma_{\textrm{s}}} = \gamma_{\textrm{s}} \sqrt{r}$. Consequently, the relations $\gamma_{\textrm{k}} \ll J \ll \gamma_{\textrm{s}}$ are guaranteed and determined by the small fraction $r\ll1$ of the metastable-state population.

To highlight the memory mechanism, and in particular the emergence of a decoherence-free subspace, we first present the interplay between $\hat{\mathcal{S}}$ and $\hat{\mathcal{K}}$ in the dark (absent the control and signal fields). Taking $\xi=-1$ in Eqs.~(\ref{eq:P_coherent_dynamics}-\ref{eq:K_coherent_dynamics}) yields the coupled-spins dynamics \begin{equation} \label{eq:dark_state_dynamics}\partial_t \begin{pmatrix} \hat{\mathcal{S}} \\ \hat{\mathcal{K}} \end{pmatrix} = \begin{pmatrix} -\gamma_{\textnormal{s}} & -i\sqrt{\gamma_{\textnormal{s}}\gamma_{\textnormal{k}}} \\ i\sqrt{\gamma_{\textnormal{s}}\gamma_{\textnormal{k}}} & -\gamma_{\textnormal{k}} \end{pmatrix} \begin{pmatrix} \hat{\mathcal{S}} \\ \hat{\mathcal{K}} \end{pmatrix}.\end{equation}
We now define the operators $\hat{\mathcal{S}}_{r}=(\hat{\mathcal{S}}+i\sqrt{r}\hat{\mathcal{K}})/\sqrt{1+r}$ and $\hat{\mathcal{K}}_{r}=(\hat{\mathcal{K}}+i\sqrt{r}\hat{\mathcal{S}})/\sqrt{1+r}$ dressed by the metastability-exchange interaction, which form an alternative set of bosonic operators\footnote{The collective spin operators are cast as bosonic operators within the Holstein-Primakoff approximation; see, \textit{e.g.}, \cite{Polzik-RMP-2010,Firstenberg-Weak-collisions}.} and  are eigenmodes of the matrix in Eq.~(\ref{eq:dark_state_dynamics}). These operators satisfy $[\hat{\mathcal{S}}_r,\hat{\mathcal{S}}_r^\dagger] = [\hat{\mathcal{K}}_r,\hat{\mathcal{K}}_r^\dagger] =1$ and $[\hat{\mathcal{S}}_r,\hat{\mathcal{K}}_r] = [\hat{\mathcal{S}}_r,\hat{\mathcal{K}}_r^\dagger] =0$ and preserve the total number of excitations $\hat{\mathcal{S}}_r^\dagger \hat{\mathcal{S}}_r + \hat{\mathcal{K}}_r^\dagger \hat{\mathcal{K}}_r = \hat{\mathcal{S}}^\dagger \hat{\mathcal{S}} + \hat{\mathcal{K}}^\dagger \hat{\mathcal{K}}$, but, through Eq.~(\ref{eq:dark_state_dynamics}), decay at different rates: $\hat{\mathcal{S}}_r$ relaxes quickly at a rate $\gamma_{\textnormal{s}}+\gamma_{\textnormal{k}}=(1+r)\gamma_{\textnormal{s}}$, while $\hat{\mathcal{K}}_r$ does not decay. Therefore, the Fock space spanned by $\hat{\mathcal{K}}_r$ can be considered as a decoherence-free subspace suitable for a  long-lived quantum memory. 

In the presence of light, collective spin excitations of alkali-metal or noble-gas spins can be coherently exchanged with photons of the input and output signal fields, but the total number of excitations decays through atomic relaxation. Carrying a derivation similar to Ref.~\cite{noble-gas-PRA}, we find that the loss of excitations is governed by
\begin{align}
    \partial_{t} & \left(\langle\hat{\mathcal{P}}^{\dagger}\hat{\mathcal{P}}\rangle+\langle\hat{\mathcal{S}}_r^{\dagger}\hat{\mathcal{S}}_r\rangle+\langle\hat{\mathcal{K}}_r^{\dagger}\hat{\mathcal{K}}_r\rangle\right)+\langle\hat{\mathcal{E}}_{\textnormal{out}}^{\dagger}\hat{\mathcal{E}}_{\textnormal{out}}\rangle-\langle\hat{\mathcal{E}}_{\textnormal{in}}^{\dagger}\hat{\mathcal{E}}_{\textnormal{in}}\rangle \nonumber \\
    & = -2\gamma_{\textnormal{p}}\langle\hat{\mathcal{P}}^{\dagger}\hat{\mathcal{P}}\rangle -2(\gamma_{\textnormal{s}}+\gamma_{\textnormal{k}})\langle \hat{\mathcal{S}}_r^\dagger\hat{\mathcal{S}}_r\rangle,
    \label{eq:MSNG_excitations_conservation_equation}
\end{align}
demonstrating that memory relaxation can be mitigated by maintaining $\langle\hat{\mathcal{P}}^{\dagger}\hat{\mathcal{P}}\rangle,\langle \hat{\mathcal{S}}_r^\dagger\hat{\mathcal{S}}_r\rangle\ll1$ small during the memory operation.  

The above analysis motivates consideration of a direct adiabatic mapping between $\hat{\mathcal{E}}_{\text{in}}$
and $\hat{\mathcal{K}}_r$ for storage, and between $\hat{\mathcal{K}}_r$ and $\hat{\mathcal{E}}_{\text{out}}$ for retrieval, maintaining  $\langle\hat{\mathcal{P}}^{\dagger}\hat{\mathcal{P}}\rangle\ll1$ and $\langle\hat{\mathcal{S}}_r^\dagger\hat{\mathcal{S}}_r\rangle\ll1$ for high memory efficiency. This mapping can be constructed by considering low-bandwidth signals with $B\ll J^2/(\gamma_\textrm{s}+\gamma_\Omega)$, where $\gamma_\Omega=|\Omega|^2/[\gamma_\textrm{p}(C+1)]$ is the  power broadening of the optical line by the control field (for $\Delta=0$), assuming that the exchange interaction rate satisfies $J\ll (\gamma_s+\gamma_{\Omega})$ 
In  Appendix~\ref{appendix:meta_stable_noble_gass_solutions}, through adiabatic elimination of Eqs.~(\ref{eq:P_coherent_dynamics})-(\ref{eq:S_coherent_dynamics}) for $\Delta=\delta_{\textrm{s}}=\delta_{\textrm{k}}=0$, we derive the dynamics and find that the evolution of the collective nuclear spins during the memory operation is governed by
\begin{equation}
    \partial_t\hat{\mathcal{K}}_r=-\dfrac{r(1+r)\gamma_{\textrm{s}}\tilde{\gamma}_{\Omega}}{\gamma_{\textrm{s}}(1+r)+\tilde{\gamma}_{\Omega}}\left(\hat{\mathcal{K}}_{r}+i\sqrt{\frac{2\eta_{C}}{r\tilde{\gamma}_{\Omega}}}\hat{\mathcal{E}}_{\textrm{in}}\right),\label{eq:K_r_dynamics}
\end{equation}
where $\tilde{\gamma}_{\Omega}\equiv\gamma_{\Omega}/(1+r)$. These equations describe the adiabatic storage and retrieval processes between the signal and the dressed noble-gas operator. During the storage stage, $\langle\hat{\mathcal{E}}_{\textnormal{in}}^{\dagger}\hat{\mathcal{E}}_{\textnormal{in}}\rangle$ is nonzero and the signal acts as a source, mapping the signal onto the collective spin for storage. During retrieval, the decay of the dressed noble-gas operator corresponds to conversion of collective spin excitations into retrieved photons (in the adiabatic limit and in the absence of other relaxation mechanisms) as shown by Eq.~(\ref{appendix:eq:MSNG_E_out}). The rate for this retrieval process, absent an input signal, is given by $\lambda=r\gamma_{\Omega}(1+r)\gamma_{\textrm{s}}/((1+r)\gamma_{\Omega}+\gamma_{\textrm{s}})$ which satisfies $\lambda<(r+r^2)\gamma_{\textrm{s}}$, highlighting the inherently low bandwidth of the retrieved signal in the adiabatic mapping. 

An analytic estimate of the storage efficiency can be derived under the assumption that the temporal profile of the optical signal is exponentially-shaped. In Appendix~\ref{appendix:meta_stable_noble_gass_solutions}, we find that the storage efficiency is then given by  
\begin{equation}
    \eta_\textrm{adiabatic}=\dfrac{C}{1+C}\left(1-\dfrac{B\gamma_\textrm{s}}{J^2(1+r)}\right),
    \label{eq:memory efficiency adiabatic}
\end{equation}
approaching unity for $C\gg1$ and $B\ll J^2/\gamma_{\textrm{s}}$ (recall that $J^2/\gamma_{\textrm{s}}=r\gamma_{\textrm{s}}$ and $r\ll1$). In Sec.~\ref{Sec:numerical},
we extend our analysis to high-bandwidth signals operating beyond the adiabatic regime.

\section{Memories based on spin-exchange collisions}\label{Sec:Spin-Exchange}
In this section, we briefly review key results of our recent analytical study on light storage in noble-gas spins via spin-exchange collisions \cite{noble-gas-PRA}, focusing on ultralow bandwidth and strong coupling configurations. This section provides background for the extensions presented in Sec.~\ref{Sec:numerical} and facilitates comparison with memories based on metastability-exchange collisions.

Nuclear spins of noble-gas atoms can couple efficiently to spins of alkali-metal atoms via the Fermi contact interaction during spin-exchange collisions \cite{Firstenberg-Weak-collisions,SEOP-Happer-RMP}.
As the spin precession angle during a single collision is very small, the total precession of the collective spin after many collisions builds up coherently, while noise and decoherence originating from the stochastic nature of the collisions remain small \cite{Firstenberg-Weak-collisions}. This different mechanism allows to realize $J$ that is greater than $\gamma_{\textrm{s}}$ and results in  strong coherent coupling between the two species as recently demonstrated \cite{Firstenberg-strong-coupling}.

In Ref.~\cite{noble-gas-PRA}, we derived Eqs.~(\ref{eq:P_coherent_dynamics}-\ref{eq:K_coherent_dynamics}) for this system with $\xi=+1$ and analyzed the dynamics in two particular limiting cases: one for low-bandwidth optical signals ($B\ll J^2/\gamma_{\textrm{s}}$), and the other for high-bandwidth signals ($B\gg \gamma_{\textrm{s}}$ and $B\gtrsim J$). The former case is analogous to the adiabatic storage strategy described above for metastability exchange, whereas the latter relies on strong coupling for efficient memory operation via sequential mapping. 

For low-bandwidth signals, we adiabatically eliminate $\hat{\mathcal{P}}$ and $\hat{\mathcal{S}}$, assuming $B\ll J^2/(\gamma_\textrm{s}+\gamma_\Omega)$.
The optical signal then couples directly to the collective noble-gas spin. Unlike with the metastability-exchange interaction, here eigenstates of $\hat{\mathcal{K}}$ span the decoherence free subspace owing to the weak nature of individual collisions; this is because the relaxation associated with the fundamental exchange process is negligible for a variety of alkali-metal and noble-gas species \cite{Firstenberg-Weak-collisions}. Notably, $\gamma_{\text{s}}$ for this configuration encompasses all relaxation mechanisms of the alkali-metal atoms\footnote{Note that, in the metastability-exchange configuration,  $\gamma_{\text{s}}$ included instead only the high relaxation rate associated with the metastability-exchange process, thus providing a lower bound for the actual relaxation.}, and the exchange interaction strength $J$ can exceed $\gamma_{\text{s}}$ \cite{Firstenberg-strong-coupling}.
To provide an estimate for the storage efficiency, we consider an exponentially-shaped pulse profile and find that, for $\gamma_{\text{k}}=0$, the storage efficiency resembles Eq.~(\ref{eq:memory efficiency adiabatic}) and is given by
\begin{equation}
    \eta_\textrm{adiabatic}=\dfrac{C}{1+C}\left(1-\dfrac{B\gamma_\textrm{s}}{J^2}\right).
    \label{eq:memory efficiency adiabatic2}
\end{equation}

\begin{figure*}[t]
\begin{centering}
\includegraphics[clip,scale=0.6]{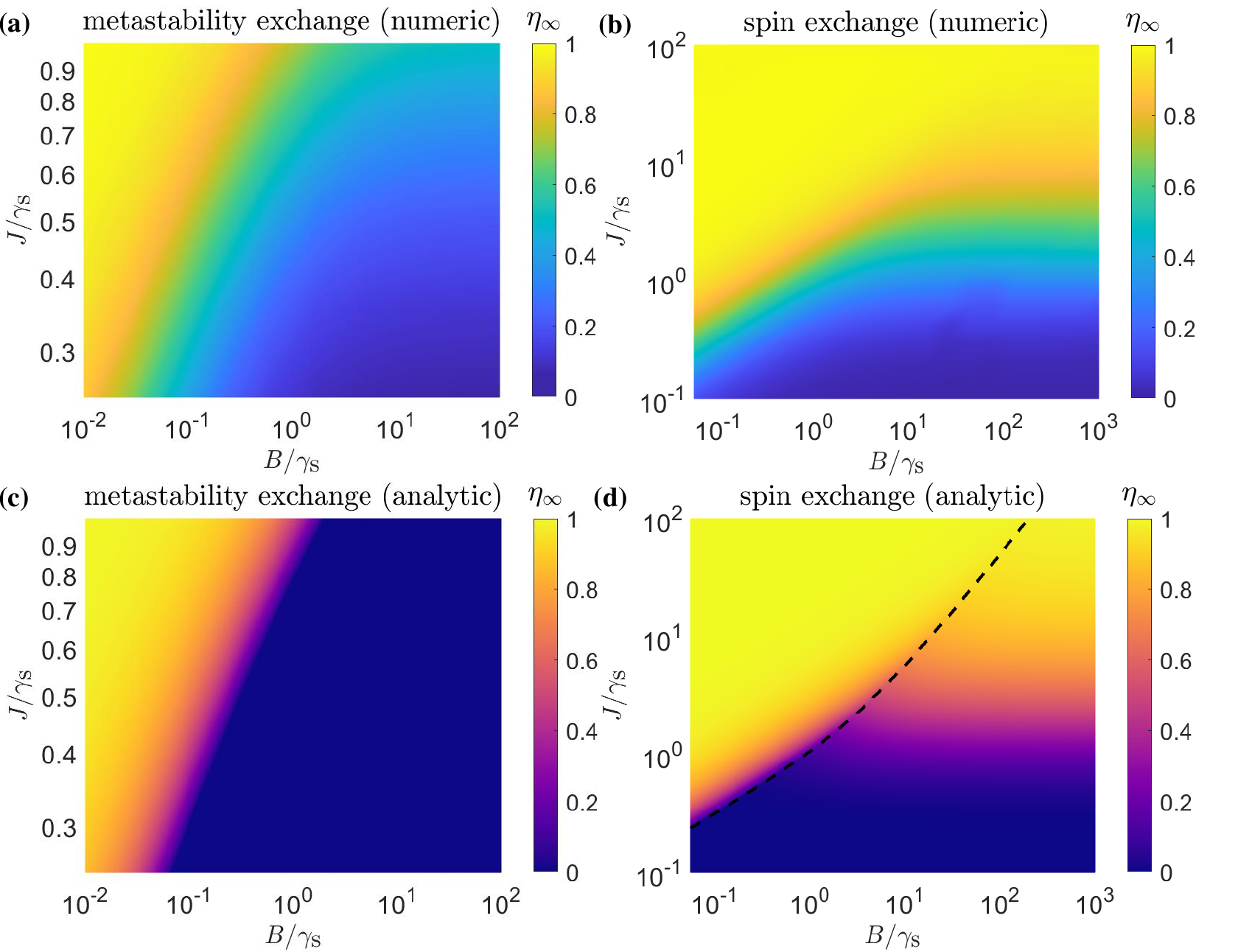}
\par\end{centering}
\centering{}\caption{\textbf{Numerically-optimized storage efficiency}.
We present the attained efficiency of the storage sequences for \textbf{(a)} metastability-exchange collisions ($\xi=-1$) and \textbf{(b)} spin-exchange collisions ($\xi=1$). The color scale shows the storage efficiency $\eta_{\infty}$ in the large-cooperativity limit. The limited range of $J/\gamma_{\text{s}}$ in \textbf{(a)} compared with \textbf{(b)} originates from the different nature of the exchange processes. 
We assume $\gamma_{\textrm{k}}=0$ and $\Delta=0$ in these calculations; see text and Appendix \ref{appendix:optimal_control_protocol_details} for details of the numerical optimization protocol.
\textbf{(c-d)} Maximal efficiency of the analytical sequences presented in Secs.~\ref{Sec:Metastable}-\ref{Sec:Spin-Exchange}, corresponding to Eq.~(\ref{eq:memory efficiency adiabatic}) in \textbf{(c)} and Eqs.~(\ref{eq:memory efficiency adiabatic2}-\ref{eq:efficiency of the fast storage scheme}) in \textbf{(d)}. 
Dashed line in \textbf{(d)} indicates the boundary at which the storage sequence change from adiabatic to sequential.
\label{fig:schemes} }\end{figure*}

For high-bandwidth signals, efficient storage is feasible when strong coupling is realized. In the limit $J \gg \gamma_\textrm{s}$, we can implement a sequential storage scheme, where light is first stored on the alkali-metal spin orientation and then mapped to the nuclear spins, implementing $\hat{\mathcal{E}}_\textrm{in}\rightarrow\hat{\mathcal{S}}(0)\rightarrow\hat{\mathcal{K}}(T')$ \cite{Firstenberg-strong-coupling}.
During the first stage, the electronic spins are excited resonantly at a rate $\gamma_\Omega=|\Omega|^2/[\gamma_\textrm{p}(C+1)]$ (similar to the operation of standard $\Lambda$-type memories \cite{Gorshkov1}) while setting $\delta_\textrm{k}\gg J$ to decouple the nuclear spins from the dynamics.
During the second stage, the electronic and nuclear spins are brought to a resonance ($\delta_\textrm{s}=\delta_\textrm{k}=0$), while $\Omega$ is turned off, allowing the two spin species to efficiently exchange their quantum state after time $T'\approx\pi/(2J)$ akin to a $\pi$-pulse of the beamsplitter Hamiltonian \cite{Polzik-RMP-2010}. For an exponentially-shaped pulse, the storage efficiency is given by \cite{noble-gas-PRA}
\begin{equation}
\eta_\text{sequential}=\frac{C}{C+1}\frac{B}{B+\gamma_\textrm{s}}\exp \left({-\frac{\pi\gamma_\textrm{s}}{2J}}\right).\label{eq:efficiency of the fast storage scheme}
\end{equation}

\section{Numerical Analysis}\label{Sec:numerical}

To extend the previous analytical results, in this section we numerically search for storage protocols that maximize the storage efficiency of the two different configurations. A similar approach, though beyond the scope of this work, can be employed to optimize the retrieval efficiency into a specific target mode.
We adopt the optimal-control tools of Ref.~\cite{Gorshkov4} and numerically solve Eqs.~(\ref{eq:P_coherent_dynamics})-(\ref{eq:K_coherent_dynamics}) for $\Delta=0$ and $\xi=\mp1$, as detailed in Appendix \ref{appendix:optimal_control_protocol_details}. We use the gradient ascent method to find the temporal profiles of $\Omega(t),\,\delta_{\textrm{s}}(t)$, and $\delta_{\textnormal{k}}(t)$ that maximize the storage efficiency $\eta=\langle\hat{\mathcal{K}}_r(s)^{\dagger}\hat{\mathcal{K}}_r(s)\rangle$ for storage based on metastability-exchange collisions and $\eta=\langle\hat{\mathcal{K}}(s)^{\dagger}\hat{\mathcal{K}}(s)\rangle$ for storage based on spin-exchange collisions for $s=T'$. 
Focusing on memory efficiency, we follow the approach in Refs.~\cite{Gorshkov1,Gorshkov2} and replace the quantum operators with the complex functions of time ($\hat{\mathcal{S}}\rightarrow\mathcal{S}$, $\hat{\mathcal{K}}\rightarrow\mathcal{K}$, and $\hat{\mathcal{E}}_{{\rm in}}\rightarrow\mathcal{E}_{{\rm in}}$), assuming the spin ensembles are initially in a vacuum state (without initial excitations). For spins, these functions represent the tilt (displacement) of the coherent spin state, and for light, they describe the field displacement for coherent states and define the temporal pulse shape. Although this method does not fully characterize the quantum state of the stored and retrieved photons, it enables the assessment of storage efficiency for general photonic signals by evaluating the output energy.

\begin{figure*}[t]
\begin{centering}
\includegraphics[clip,scale=0.46]{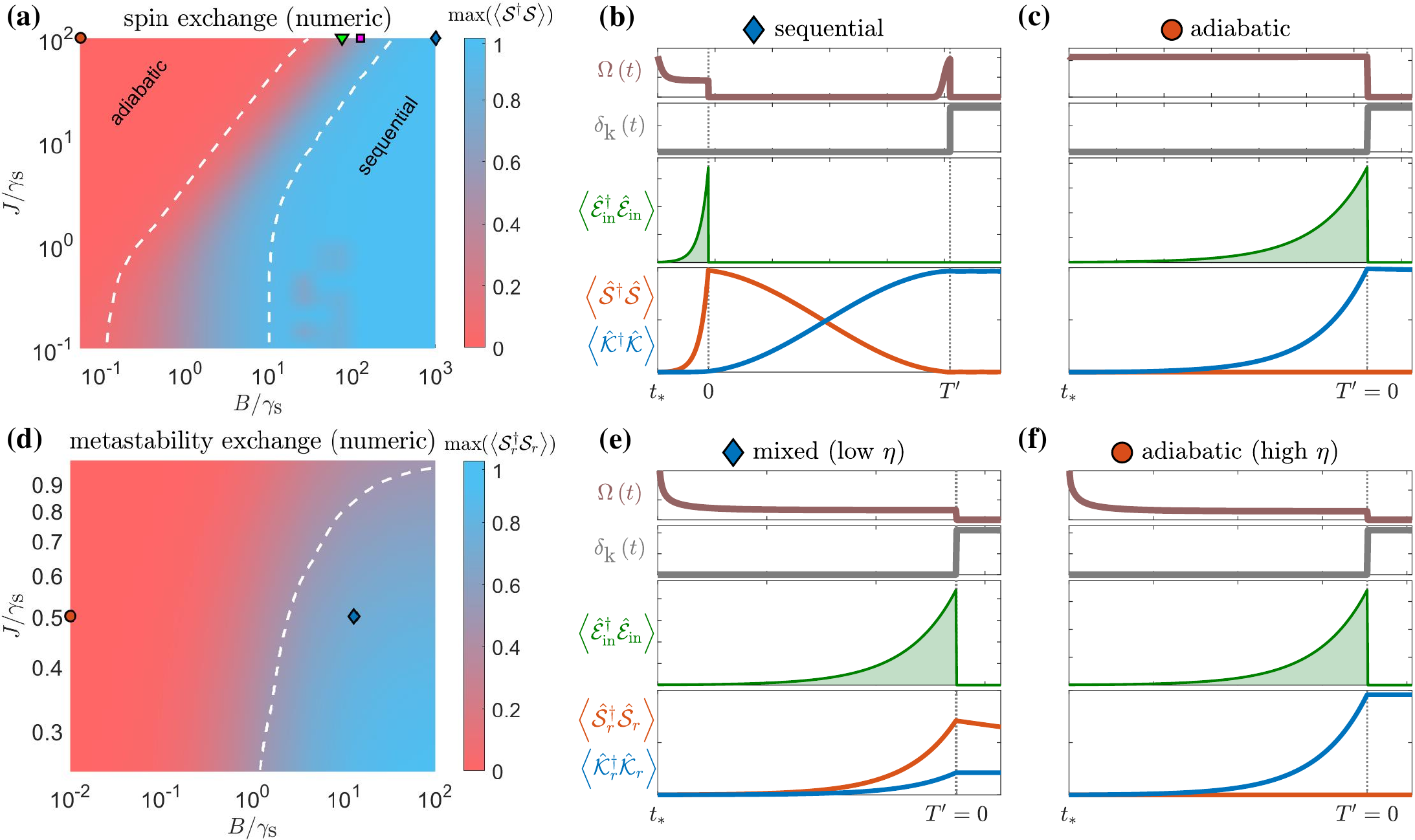}\par\end{centering}
\centering{}\caption{\textbf{Optimal storage sequences}. 
\textbf{(a), (d)} Maximal excitation $\langle\hat{\mathcal{S}}^\dagger\hat{\mathcal{S}}\rangle$ of the electron spins during
the storage sequence for the numerically optimized solutions for spin exchange collisions \textbf{(a)} and metastability exchange collisions \textbf{(d)}. %$C=100$ and $\gamma_{\textrm{k}}=0$.
Two distinct regimes of nearly complete excitation (blue) or nearly no excitation %\RSver{of the electronic spin} 
(red) 
are observed for the optimal solutions as a function of the bandwidth $B$ and coupling strength $J$. 
Dashed lines denote the condition(s) $\max(\langle\hat{\mathcal{S}}^\dagger\hat{\mathcal{S}}\rangle)=0.9$ and $\max(\langle\hat{\mathcal{S}}^\dagger\hat{\mathcal{S}}\rangle)=0.1$ in \textbf{(a)} and $\max(\langle\hat{\mathcal{S}}^\dagger\hat{\mathcal{S}}\rangle)=0.5$ in \textbf{(d)}, marking the approximate boundaries of these two regimes, qualitatively showing where the optimal solutions follow the adiabatic or sequential strategies.
\textbf{(b), (c)} Particular numerically-optimized solutions, corresponding to the blue diamond and red circle in \textbf{(a)}, indicating sequential-like and adiabatic-like storage strategies, respectively. \textbf{(e), (f)} Particular numerically-optimized solutions corresponding to the blue diamond and red circle in \textbf{(d)}, indicating solutions with different bandwidth and different electron excitation, where \textbf{(f)} follows an adiabatic solution (see Appendix~\ref{appendix:meta_stable_noble_gass_solutions}).
The optimization parameters used in \textbf{(a)} and \textbf{(d)} are identical to the ones presented in Fig.~\ref{fig:schemes}(b) and Fig.~\ref{fig:schemes}(a), respectively. We find that $\delta_{\text{s}}(t)$ (not shown) is near zero throughout the storage sequence, a result of the constant phase of the input pulse, as we discuss in Appendix \ref{sec:optimized_delta_S}. The increase in ${\Omega}$ towards the end of the storage sequence in \textbf{(b)} is merely an artifact of the optimizer trying to decouple the alkali and noble-gas spins more efficiently, as there is no constraint on the control field amplitude in the optimization, see Appendix \ref{appendix:optimal_control_protocol_details}. Intermediate bandwidth solutions corresponding to the green triangle and the magenta square in \textbf{(a)} are shown in Fig.~\ref{fig:SE_intermediate_solutions}. \label{fig:optimal_storage_strategies}}
\end{figure*}

The input pulse $\mathcal{E}_{\textrm{in}}\left(t\right)$ spans from time $t=t_* < 0$ to $t=0$. The end time of the storage sequence, $T'$, varied during the optimization process as a function of $\gamma_{\textrm{s}}$ and $J$ (see Appendix \ref{appendix:structure_of_efficiency}). We report the maximal efficiency, where $\eta(s)$ achieves its highest value for any $s\leq T'$. In Appendix \ref{appendix:structure_of_efficiency}, we show that, for negligible noble-gas relaxation ($\gamma_{\textrm{k}}=0$ for spin-exchange collisions and $\gamma_{\textrm{k}}=r\gamma_{\textrm{s}}$ for metastability-exchange collisions), the storage efficiency can be cast as  \begin{equation}\eta(C,\gamma_\textrm{s},B,J)=\frac{C}{C+1}\eta_\infty(B/\gamma_\textrm{s},J/\gamma_\textrm{s}),\label{eq:eta_infty}\end{equation}depending trivially on $C$ and non-trivially on the parameters $B/\gamma_\textrm{s}$ and $J/\gamma_\textrm{s}$. Here we study the dependence of the performance on these parameters for multiple different pulse shapes.
We fix $C=100$ in the numerical simulations and present the reduced efficiency parameter $\eta_\infty$ in Fig.~\ref{fig:schemes}(a) for storage based on metastability exchange collisions ($\xi=-1$ and $J<\gamma_\mathrm{s}$) and in Fig.~\ref{fig:schemes}(b) for storage based on spin-exchange collisions ($\xi=+1$). These calculations use an exponentially-shaped input pulse $\mathcal{E}_\textrm{in}(t) = A\sqrt{2B}\exp(Bt)$ for $t\le0$, taking  $2B$ as the pulse bandwidth, $t_*=-3/B$ and $A=e^{3}/\sqrt{e^{6}-1}$ fixing unity excitation. We note that similar results are obtained for Lorenzian- and Gaussian-shaped pulses with similar bandwidths, as we show in Appendix \ref{appendix:dependence_on_pulse_shape}.

The optimization shows that efficient storage of light is obtained for low-bandwidth signals $B<\gamma_\Omega,\,J^2/\gamma_\textrm{s}$, as expected from the adiabatic mapping. 
For storage based on spin-exchange collisions, we find that efficient storage can be realized for any $J\gg \gamma_{\text{s}}$ as expected in the strong-coupling regime. We find numerically that the optimal value of $\delta_{\text{s}}$ is near zero. We discuss the optimal value of $\delta_{\text{s}}$ and its dependence on the signal phase in Appendix \ref{sec:optimized_delta_S}

\begin{table*}[t]
\centering{}
\resizebox{\textwidth}{!}{\begin{tabular}{|c|c|c|c|c|c|c|c|}
\hline
Platform & $\gamma_{\mathrm{s}}$ [$(2\pi)$ Hz] & $J$ [$(2\pi)$ Hz] & $J/\gamma_{\mathrm{s}}$ & $C$ & 1/$\gamma_{\text{k}}$ & Time-bandwidth & Efficiency \\
& & & & & & product $B/\gamma_{\text{k}}$& $\eta_{\textrm{tot}}$ \\
\hline
\hline
Metastable and ground & $5\cdot 10^{6}$ \cite{Sinatra-2005-Metastable} & $5\cdot 10^{3}$ \cite{Sinatra-2005-Metastable} & $10^{-3}$ & $500$ \cite{Sinatra-2005-Metastable} & $0.2^\dagger$ s \cite{Sinatra-2005-Metastable}  $100^*$ h \cite{Heil-Noble-T2} & $\sim ~10^5$ & 76\% | adiabatic | $B\sim 0.1\;(2\pi) \textrm{Hz}$ \\
state helium-3 & & & & & & & \\
\hline
Alkali noble-gas & $6-15$ \cite{noble-gas-PRA,Budker-Romalis-Magnetometry} & $490-690$ \cite{Firstenberg-strong-coupling,noble-gas-PRA} & $33-115$ & $37$ \cite{noble-gas-PRA} & $2^\dagger$ h \cite{Firstenberg-strong-coupling} $100^*$ h \cite{Heil-Noble-T2} & $\gtrsim10^{10}$ & 94\% | adiabatic | $B\lesssim 10\;(2\pi) \textrm{Hz}$ \\
mixture & & & & & & & 89\% | sequential | $B\gtrsim 1\;(2\pi) \textrm{MHz}$ \\
\hline
\end{tabular}}\caption{\textbf{Estimation of feasible experimental parameters}. The parameters are adapted from the provided references, and the  efficiencies are computed from Eqs.~(\ref{eq:memory efficiency adiabatic}-\ref{eq:efficiency of the fast storage scheme}), for the second configuration taking the maximal values of the parameter range and estimating the total memory efficiency using  $\eta_{\text{tot}}\approx\eta^2$. $\dagger,*$ denote configurations in which the exchange interaction is turned on or off, respectively. It is assumed that, during the memory time, following the storage, the exchange interaction is turned off; see text.
\label{tab:experimental_configs}}
\end{table*}

We compare the numerically optimized efficiency with the one associated with the simple analytical schemes in Sec.~\ref{Sec:Metastable}-\ref{Sec:Spin-Exchange}, whose efficiencies are given in Eqs.~(\ref{eq:memory efficiency adiabatic}-\ref{eq:efficiency of the fast storage scheme}). In Fig.~\ref{fig:schemes}(c), we present the positive values of the storage efficiency in Eq.~(\ref{eq:memory efficiency adiabatic}) for storage based on metastability-exchange collisions, and in Fig.~\ref{fig:schemes}(d) the maximal storage efficiency of Eqs.~(\ref{eq:memory efficiency adiabatic2}-\ref{eq:efficiency of the fast storage scheme}) for storage based on spin-exchange collisions. Interestingly, the analytic expressions for the spin-exchange configuration provide a good approximation to the numerically-calculated efficiency even away from their validity limits, especially for the spin-exchange case. For metastability-exchange collisions, on the other hand, the numerical solution is similar to the analytic solution at low pulse bandwidths, but displays remarkably high efficiencies at large bandwidth far from the adiabatic regime, owing to a better performance of optimal non-adiabatic solutions. Similar behavior has been observed in Ref.~\cite{Gorshkov4}, where numerically-optimized solutions were shown to dramatically increase the memory bandwidth.

The similarity between the analytical and numerical efficiency maps suggests that the numerically optimized solutions may resemble the analytical ones. 
To explore this, in Fig.~\ref{fig:optimal_storage_strategies}(a), we present the maximal excitation of the electronic spin at intermediate times during the storage process, $\text{max}(\langle\hat{\mathcal{S}}^\dagger(t)\hat{\mathcal{S}}(t)\rangle)$  for spin-exchange collisions ($\xi=1$). Strikingly, we find that, in most of the parameter space, the collective electron spin is either nearly unexcited or temporarily holds a significant portion of the input excitations from the signal field, which are mapped directly onto the electrons. Dashed white lines indicate the $10\%$ and $90\%$ excitation values, respectively. In Fig.~\ref{fig:optimal_storage_strategies}(b-c), we plot two of the numerically-optimized solutions, representing the blue diamond and red circle symbols in Fig.~\ref{fig:optimal_storage_strategies}(a), respectively. In Fig.~\ref{fig:optimal_storage_strategies}(b), we find that the numerically optimized solution is nearly identical to the sequential scheme, where the excitation is first mapped onto the alkali-metal spins, and only later mapped to the noble-gas spins in the absence of a control field. In Fig.~\ref{fig:optimal_storage_strategies}(c), the numerically-optimized solution is similar to the adiabatic scheme, where the control field is constantly on, and the electron spins are nearly unexcited. The solutions between the two dashed lines in Fig.~\ref{fig:optimal_storage_strategies}(a) belong to neither of these schemes, yet could attain high efficiency. Examples of these solutions are plotted in Fig.~\ref{fig:SE_intermediate_solutions}.

We repeat the preceding analysis for the $\xi=-1$ case, as shown in Fig.~\ref{fig:optimal_storage_strategies}(d-f). For low-bandwidth signals, we find that numerically optimized solutions attain the form of the adiabatic scheme. For high-bandwidth signals, we find that the solutions are non-adiabatic and involve increased electronic spin population, yet show moderate efficiency owing to the restrictive ratio of $J/\gamma_{\text{s}}$ that can be physically realized using the metastability-exchange mechanism.

\section{Discussion}\label{Sec:summary}

The storage efficiencies of noble-gas spins based on metastability-exchange collisions or spin-exchange collisions can theoretically approach unity for large cooperativity ($C\gg1$) and sufficiently low pulse bandwidth $B\ll \gamma_{\text{s}}$. In the limit of zero bandwidth ($B\ll J^2/\gamma_{\text{s}}$), the storage efficiency of the adiabatic scheme approaches $C/(C+1)$, which is the maximal efficiency of optically accessible $\Lambda$-type memories. For these scenarios, we find that the numerically optimal sequences follow adiabatic-like protocols, where the electron spins remain nearly unexcited,  and the optical control field $\Omega$ is kept high throughout the pulse.

For higher pulse bandwidth  ($B\gtrsim\gamma_{\text{s}}$), it is optimal to transfer a significant part or even most of the excitation to the electronic spins first, and then reduce the power of the optical control field during the transfer of excitation from the electronic spins to noble-gas spins. While configurations based on spin-exchange collisions can remain efficient if the exchange coupling is sufficiently large ($J\gtrsim\gamma_{\text{s}}$), the efficiency of metastability-exchange configurations is more limited owing to the physical limitation on the exchange-rate strength relative to the accompanying relaxation. 

Both configurations examined in this work are experimentally feasible. 
Control of Rabi frequency $\Omega$ and detuning $\delta_{\textrm{s}}$ is implemented by tuning the control laser power and frequency, respectively, whereas the difference 
$\delta_{\textrm{k}}-\delta_{\textrm{s}}$ can be controlled by an external magnetic field owing to the different magnetic dipole moments (gyromagnetic ratios) of the electron and nuclear spins \cite{noble-gas-PRA,Sinatra-2005-Metastable}.
For the metastability-exchange-based configuration, using the parameters from Ref.~\cite{Sinatra-2005-Metastable} and taking $r=10^{-6}$, storage of low-bandwidth signals can be implemented efficiently. We estimate a typical bandwidth for this scheme to be limited by $B\lesssim (2\pi)\,1\,$Hz, \textit{e.g.}~with efficiency $\eta_{\textrm{tot}}\approx0.76$ for a bandwidth of $(2\pi)\,0.1$ Hz, where  $\eta_{\textrm{tot}}\approx\eta^{2}$ is the approximate combined memory efficiency of the storage and retrieval stages \cite{Gorshkov1,noble-gas-PRA}. For the spin-exchange-based configuration, we consider a mixture of potassium and helium-3. 
For high-bandwidth pulses, we estimate $B \gtrsim (2\pi)$ 1 MHz and efficiency of $\eta_{\textrm{tot}}\approx 0.89$ using the sequential strategy, whereas, for very low-bandwidth pulses ($B\lesssim (2\pi) 10$ Hz), even higher efficiencies $\eta_{\textrm{tot}}\approx 0.94$ can be realized with the adiabatic scheme. We summarize these results in Table.~\ref{tab:experimental_configs}. We assume that, during the long memory time, the relaxation of the noble-gas spins by coupling to the electronic spins is suppressed. In metastability-exchange collisions, this can be implemented by turning off the discharge, which leaves the entire noble-gas population in the electronic ground state manifold (with the optical signal mapped onto the spin states). For spin-exchange collisions, turning off can be realized by first turning off the optical pumping beam, which leaves only spin-rotation coupling. The latter can be then suppressed by cooling the cell and reducing the alkali-atom vapor density through condensation. Owing to the large separation of scales between  $\gamma_{\textrm{k}}$ and $\gamma_{\textrm{s}}$, we set the former to zero in all the calculations performed in this work.

It is interesting to compare our results with the optimal storage strategy for electron spins in a $\Lambda$-type system \cite{Gorshkov1}.
The latter exhibits adiabatic-like optimal solutions, achieved by shaping the control fields over time to support high-bandwidth pulses. Since the exchange interaction strength is fixed and cannot be modulated over time to match the input pulse shape, the optimal strategy deviates from the adiabatic  scheme for higher pulse bandwidths. In this case, the excitations are temporarily stored- either fully or partially- on the electronic spins, in contrast to the adiabatic strategy, where the electronic spins remain only weakly excited. The strong coupling regime is particularly promising from an application perspective, offering an avenue to realize high-time-bandwidth-product memories. The time-bandwidth product is one of the main figures of merit for optical quantum memories \cite{Tittel-review-nature-photonics}, defined as the product of the pulse bandwidth and the memory coherence time. For noble gas spins, the coherence time can be exceptionally long ($1/\gamma_{\text{k}}\gtrsim 1$ hour), underscoring the remarkable potential of this technology.

\begin{acknowledgments}
ER, RS, EP, and OF\ acknowledge financial support by the Israel Science Foundation, the US-Israel BSF and US NSF, the Pazy Foundation, the Minerva Foundation with funding from the Federal German Ministry for Education and Research, the Estate of Louise Yasgour, and the Laboratory in Memory of Leon and Blacky Broder. AVG was  supported in part by DARPA SAVaNT ADVENT, ARO MURI, AFOSR MURI, AFOSR, DoE ASCR Accelerated Research in Quantum Computing program (award No.~DE-SC0020312), NSF STAQ program, DoE ASCR Quantum Testbed Pathfinder program (awards No.~DE-SC0019040 and No.~DE-SC0024220), and NSF QLCI (award No.~OMA-2120757). Support is also acknowledged from the U.S.~Department of Energy, Office of Science, National Quantum Information Science Research Centers, Quantum Systems Accelerator. 
\end{acknowledgments}

\clearpage
\appendix
\onecolumngrid 

\section{Metastability-exchange equations of motion}\label{appendix:meta_stable_noble_gasses}
In this appendix, we describe the correspondence between the notation used for the dynamics of collective operators in Ref.~\cite{Sinatra-2005-Metastable} and the notation used in Eqs.~(\ref{eq:P_coherent_dynamics}-\ref{eq:K_coherent_dynamics}).

In Ref.~\cite{Sinatra-2005-Metastable}, the dynamics of the proposed metastability-exchage based memory is described in Eqs.~(2-5) therein. We can relate the notation in Ref.~\cite{Sinatra-2005-Metastable} to our notation as follows: the field $\hat{\mathcal{E}}_{\textrm{in}}\equiv \hat{A}_{\textrm{in}}$ and coherences $\hat{\mathcal{P}}\equiv -\hat{S}_{23}/\sqrt{n}$, $\hat{\mathcal{S}}\equiv \hat{S}_{21}/\sqrt{n}$, $\hat{\mathcal{K}}\equiv i\hat{I}_{09}/\sqrt{N}$. Additionally, the rates map as $\gamma_{\rm p}\equiv\gamma$, $\gamma_{\rm s}\equiv\gamma_{\rm m}$, $\gamma_{\rm k}\equiv\gamma_{\rm f}$, $\delta_{\rm s}\equiv-\delta$ and $\delta_{\rm k}\equiv-\delta_{I}$, where $n$ and $N$ denote the number densities in the metastable and ground state, respectively, using the notation from Ref.~\cite{Sinatra-2005-Metastable} (not to be confused with $N$ representing the number of photons in the pulse in the rest of this work). We also define $J\equiv \sqrt{\gamma_{\textnormal{s}}\gamma_{\textnormal{k}}}$ and assume $\Omega$ is real. Notably, unlike the model in Sec.~\ref{Sec:Spin-Exchange}, which includes, and is limited by, relaxation processes whose mechanism is unrelated to the underlying exchange interaction, here the metastability-exchange interaction itself generates nonzero relaxation. Adopting the assumptions in Ref.~\cite{Sinatra-2005-Metastable}, we consider only these relaxation rates in this model and neglect all other spin relaxation mechanisms that are not associated with the metastability-exchange interaction.

\section{Adiabatic storage sequence for memories based on metastability-exchange collisions}\label{appendix:meta_stable_noble_gass_solutions}

In this appendix, we analytically study a storage sequence that relies on adiabatic following of the excitation from light to noble-gas excitations that is suitable for low-bandwidth pulses, similar to the adiabatic sequence derived in Ref.~\cite{noble-gas-PRA}. Assuming that $\hat{\mathcal{P}}$ follows adiabatically the input pulse, it can be expressed as 
\begin{equation}\hat{\mathcal{P}}=\frac{i\Omega\hat{\mathcal{S}}+i\sqrt{2\gamma_{\textnormal{p}}C}\hat{\mathcal{E}}_{\text{in}}}{\gamma_{\textnormal{p}}(1+C)+i\Delta}.\label{eq:P_adiab}\end{equation}
Rewriting Eqs.~(\ref{eq:S_coherent_dynamics}-\ref{eq:K_coherent_dynamics}) using the dressed operators  $\hat{\mathcal{S}}_{r}$ and $\hat{\mathcal{K}}_{r}$, we find
\begin{align}
    \partial_{t}\hat{\mathcal{S}}_{r}&=-\left[\gamma_{\textrm{s}}(1+r)+\tilde{\gamma}_{\Omega}+i\tilde{\delta}_{\textrm{s}}+ir\tilde{\delta}_{\textrm{k}}\right]\hat{\mathcal{S}}_{r}+\sqrt{r}\left(\tilde{\delta}_{\textrm{k}}-\tilde{\delta}_{\textrm{s}}+i\tilde{\gamma}_{\Omega}\right)\hat{\mathcal{K}}_{r}-\sqrt{2\tilde{\gamma}_{\Omega}}\sqrt{\eta_{C}}e^{-i\arg{\Omega}}\hat{\mathcal{E}}_{\textrm{in}},
    \label{appendix:eq:MSNG_S_r}
    \\
    \partial_{t}\hat{\mathcal{K}}_{r}&=-\left(r\tilde{\gamma}_{\Omega}+i\tilde{\delta}_{\textrm{k}}+ir\tilde{\delta}_{\textrm{s}}\right)\hat{\mathcal{K}}_{r}-\sqrt{r}\left(\tilde{\delta}_{\textrm{k}}-\tilde{\delta}_{\textrm{s}}+i\gamma_{\Omega}\right)\hat{\mathcal{S}}_{r}-i\sqrt{r}\sqrt{2\tilde{\gamma}_{\Omega}}\sqrt{\eta_{C}}e^{-i\arg{\Omega}}\hat{\mathcal{E}}_{\textrm{in}},
    \label{appendix:eq:MSNG_K_r}
\end{align}
where all variables with tilde are rescaled by $(1+r)$, \textit{e.g.} $\tilde{\gamma}_{\Omega}\equiv\gamma_{\Omega}/(1+r)$. Adiabatically eliminating $\hat{\mathcal{S}}_{r}$ in Eq.~(\ref{appendix:eq:MSNG_S_r}), substituting  the resulting  $\hat{\mathcal{S}}_{r}$ into Eq.~(\ref{appendix:eq:MSNG_S_r}), and taking $\Delta=\delta_{\textrm{s}}=\delta_{\textrm{k}}=0$ yields Eq.~(\ref{eq:K_r_dynamics}). Furthermore, we can derive an expression for the output field $\hat{\mathcal{E}}_{\textrm{out}}=\hat{\mathcal{E}}_{\textrm{in}}+i\sqrt{2\gamma_{\textrm{p}}}\hat{\mathcal{P}}$ in this limit:
\begin{equation}
\hat{\mathcal{E}}_{\textrm{out}}=\left[1-2\eta_{C}\frac{\gamma_{\textrm{s}}(1+r)}{\gamma_{\textrm{s}}(1+r)+\tilde{\gamma}_{\Omega}}\right]\hat{\mathcal{E}}_{\textrm{in}}+i\sqrt{\frac{2\eta_{C}}{r\tilde{\gamma}_{\Omega}}}\frac{r(1+r)\gamma_{\textrm{s}}\tilde{\gamma}_{\Omega}}{\gamma_{\textrm{s}}(1+r)+\tilde{\gamma}_{\Omega}}e^{-i\arg{\Omega}}\hat{\mathcal{K}}_{r}.
\label{appendix:eq:MSNG_E_out}
\end{equation}

To solve Eq.~(\ref{eq:K_r_dynamics}), we note that the solution for a general equation of the form \begin{equation}\partial_{t}\hat{\mathcal{K}}_{r}=-\gamma\hat{\mathcal{K}}_{r}+A\hat{\mathcal{E}}_{\textrm{in}}\end{equation} is given by \begin{equation}\hat{\mathcal{K}}_{r}\left(0\right)=\int_{-\infty}^{0}Ae^{-\int_{t}^{0}\gamma\textrm{d}s}\hat{\mathcal{E}}_{\textrm{in}}\textrm{d}t,\end{equation} where we can identify the parameter $\eta'=\left|A\right|^{2}/(2\gamma)$ and the transfer function $h_{r}(t)=Ae^{-\int_{t}^{0}\gamma\textrm{d}s}$ satisfying $\int_{-\infty}^{0}\frac{\left|h_{r}\right|^2}{\eta'}\textrm{d}t\leq 1$. In our case then, we find
\begin{equation}
\eta'(t)=\frac{C}{C+1}\dfrac{\gamma_{\textrm{s}}(1+r)}{\gamma_{\textrm{s}}(1+r)+\tilde{\gamma}_{\Omega}(t)}.
    \label{appendix:eq:MSNG_K_r_efficiency}
\end{equation}
For an exponentially-shaped input of bandwidth $2B$, optimal storage follows if $\gamma=B$ \cite{noble-gas-PRA}, which can be realized by a constant-amplitude control field \begin{equation}\label{eq:gammaOmega}\tilde{\gamma}_{\Omega}=\frac{B\gamma_{\textrm{s}}(1+r)}{\gamma_{\textrm{s}}r(1+r)-B}.\end{equation} Under these conditions, the storage efficiency of this pulse becomes $\eta=\eta'$, and substitution of Eq.~(\ref{eq:gammaOmega}) into Eq.~(\ref{appendix:eq:MSNG_K_r_efficiency}) yields Eq.~(\ref{eq:memory efficiency adiabatic}).
\section{Numerical Optimization}\label{appendix:optimal_control_protocol_details}
In this appendix, we describe the numerical optimization protocol used in this work. We focuses on the regime  $\gamma_{\textnormal{p}}C\gg B$, where the optical dipole $\hat{\mathcal{P}}$ can be adiabatically-eliminated. This approximation simplifies Eqs.~(\ref{eq:P_coherent_dynamics}-\ref{eq:K_coherent_dynamics}) and yields a coupled set of equations describing the collective spins:
\begin{align}
\partial_{t}\mathcal{S} &=-(\gamma_{\text{s}}+\Gamma_{\Omega}+i\delta_{\text{s}})\mathcal{S}-iJ\mathcal{K}-Q\Omega^{*}\mathcal{E}_{\text{in}},\label{eq:numerical_equations_alkali}\\
\partial_{t}\mathcal{K} &=-(\gamma_{\textnormal{k}}+i\delta_{\text{k}})\mathcal{K}-i\xi J\mathcal{S}.\label{eq:numerical_equations_noble}
\end{align}
Here, we use $\Gamma_{\Omega}=\left|\Omega\right|^{2}/(\gamma_{\textrm{p}}\left(1+C\right)+i\Delta)$ and $Q=\sqrt{2\gamma_{\textrm{p}}C}/(\gamma_{\textrm{p}}\left(1+C\right)+i\Delta)$, and we exchanged the operators with complex functions as detailed Sec.~\ref{Sec:numerical}. We set $\Delta=0$,
assume that $\Omega$ is real, and allow the simulation to optimize the control fields with storage end time %for a duration of 
$T'=\pi/[2\text{max}(\sqrt{J^2-\gamma_{\text{s}}^2/4},\gamma_{\text{s}})]$ to enable operation in the sequential scheme, which requires such duration \cite{noble-gas-PRA}. For the spin-exchange configuration ($\xi=1$), we assume $\gamma_{\textnormal{k}}=0$, and, for the metastability-exchange configuration ($\xi=-1$), we assume the minimal value $\gamma_{\textnormal{k}}=r\gamma_{\textnormal{s}}$, which results in zero relaxation of the operator $\hat{\mathcal{K}}_r$. Because the temporal profile of the signal starts from some negative time $t_*$ and ends at time $t=0$, taking $T'\geq0$ enables an additional interaction between the spins $\mathcal{S}$ and $\mathcal{K}$ after the end of the incoming optical signal.

We generalize the variational technique introduced in Ref.~\citep{Gorshkov4}
to find numerically the optimal control pulses for $\Omega(t),\,\delta_{\text{s}}(t)$,
and $\delta_{\text{k}}(t)$ that maximize the storage efficiency given $B/\gamma_{\textrm{s}}$ and $J/\gamma_{\textrm{s}}$.
We use the gradient ascent method to vary the control functions between different iterations following Ref.~\citep{Gorshkov4}, aiming to maximize the functional
\begin{equation}
\begin{aligned}\Phi & =\Phi_{0}-\frac{1}{2}\int_{-\infty}^{T'}{\rm d}t\Bigl[k^{\ast}\left(\partial_{t}\mathcal{K}+(\gamma_{\textnormal{k}}+i\delta_{\text{k}})\mathcal{K}+i\xi J\mathcal{S}\right)\\
+ & s^{\ast}\left(\partial_{t}\mathcal{S}+(\gamma_{\textnormal{s}}+\gamma_{\Omega}+i\delta_{\text{s}})\mathcal{S}+iJ\mathcal{K}+Q\Omega\mathcal{E}_{{\rm in}}\right)+{\rm h.c.}\Bigr],
\end{aligned}
\label{eq: cost function}
\end{equation}
where $\Phi_{0}=\frac{1}{2}|\mathcal{K}_{r}(T')|^{2}$ for metastability-exchange collisions ($\xi=-1$) and $\Phi_{0}=\frac{1}{2}|\mathcal{K}(T')|^{2}$ for spin-exchange collisions ($\xi=+1$). The functional $\Phi$ describes the number of stored excitations
onto the collective long-lived spin operator, whereas $s(t)$ and
$k(t)$ are the Lagrange multipliers,  which ensure that Eqs.~(\ref{eq:numerical_equations_alkali}-\ref{eq:numerical_equations_noble}) for $\mathcal{S}$ and $\mathcal{K}$ are satisfied.  The variations
in $\Phi$ with respect to the relevant functions vanish for the optimal
solution. Variational calculus yields the equations of motion
for the Lagrange multipliers:
\begin{align}
\partial_{t}s & =(\gamma_{\textnormal{s}}+\gamma_{\Omega}-i\delta_{\text{s}})s-i\xi Jk,\label{eq:s-lagrange}\\
\partial_{t}k & =(\gamma_{\textnormal{k}}-i\delta_{\text{k}})k-iJs,\label{eq:k-lagrange}
\end{align}
with the conditions $s(T')=0$ and $k(T')=\mathcal{K}(T')$ for $\xi=+1$ and $s(T')=-i\sqrt{\frac{r}{1+r}}\mathcal{K}_{r}(T')$ and $k(T')=\frac{1}{\sqrt{1+r}}\mathcal{K}_{r}(T')$ for $\xi=-1$.
Note that we are interested in computing the multipliers $s(t)$ and
$k(t)$ for $t\leq T'$, and therefore we numerically
solve Eqs.~(\ref{eq:s-lagrange}-\ref{eq:k-lagrange}) backwards
in time, from $t=T'$. In every iteration, we first solve
Eqs.~(\ref{eq:numerical_equations_alkali}-\ref{eq:numerical_equations_noble}),
compute $\mathcal{K}(T)$, and then solve Eqs.~(\ref{eq:s-lagrange}-\ref{eq:k-lagrange}).
We use these solutions to calculate the functional derivatives of the
control functions $\Omega(t),\,\delta_{\text{s}}(t)$, and $\delta_{\text{k}}(t)$:
\begin{align}
\frac{\partial\Phi}{\partial\tilde{\Omega}} & =-2\tilde{\Omega}\text{re}(s^{\ast}\mathcal{S})-\sqrt{\frac{2C}{C+1}}\text{re}(s)\mathcal{E}_{{\rm in}},\\
\frac{\partial\Phi}{\partial\delta_{\text{s}}} & ={\rm im}(s^{\ast}\mathcal{S}),\\
\frac{\partial\Phi}{\partial\delta_{\text{k}}} & ={\rm im}(k^{\ast}\mathcal{K}),
\end{align}
where we use the normalized rate $\tilde{\Omega}=\sqrt{\gamma_{\Omega}}$.
This set of equations is used to update the control functions for
the next iteration. The control functions in the $n^{\text{th}}$
iteration are determined using the gradient ascent method with momentum 
\citep{gradient-ascent-1,gradient-ascent-2}:
\begin{align}
\tilde{\Omega}^{\left(n\right)} & =(1+\alpha_{n})\tilde{\Omega}^{\left(n-1\right)}-\alpha_{n}\tilde{\Omega}^{\left(n-2\right)}+\frac{1}{\lambda_{\tilde \Omega}}\frac{\partial\Phi}{\partial\tilde{\Omega}},\\
\delta_{\text{s}}^{\left(n\right)} & =(1+\alpha_{n})\delta_{\text{s}}^{\left(n-1\right)}-\alpha_{n}\delta_{\text{s}}^{\left(n-2\right)}+\frac{1}{\lambda_{\delta_{\text{s}}}}\frac{\partial\Phi}{\partial\delta_{\text{s}}},\\
\delta_{\text{k}}^{\left(n\right)} & =(1+\alpha_{n})\delta_{\text{k}}^{\left(n-1\right)}-\alpha_{n}\delta_{\text{k}}^{\left(n-2\right)}+\frac{1}{\lambda_{\delta_{\text{k}}}}\frac{\partial\Phi}{\partial\delta_{\text{k}}}.
\end{align}
Here we choose $\alpha_{n}=0.9$ for $n\geq3$ and $\alpha=0$ otherwise and use $\lambda$s to denote the inverse step size between iterations.

For the numerical results, we assume the minimal possible values of $\gamma_{\text{k}}$ for these two approaches, $\gamma_{\text{k}}=r\gamma_{\text{s}}$ and $\gamma_{\text{k}}=0$ for metastability-exchange and spin-exchange collisions, respectively. In the main text, we choose the input signal to have an exponential temporal profile of the form
\begin{equation}
\mathcal{E}_{{\rm in}}(t)=A\sqrt{\frac{2}{T}}e^{t/T},\,\,\,\,\,\,\,\:\,\,t_{\star}\leq t\leq 0,\label{eq:exponentially shaped pulse}
\end{equation}
and zero otherwise, setting $T=B^{-1}$. For $\Lambda$-type memories in the adiabatic regime, optimal storage of an exponentially-shaped signal (with $t_{\star}=-\infty$) is done with a square control pulse \citep{Gorshkov1}. Therefore, this choice allows for a direct comparison to simple analytic expressions for the efficiencies. We numerically truncate the pulse at $t_{\star}=-3T$ to reduce the numerical complexity. In this case, setting $A=e^3/\sqrt{e^{6}-1}$ guarantees the normalization
$\int_{-3T}^{0}|\mathcal{E}_{{\rm in}}(t)|^{2}dt=1$. The overlap $\mathcal{F}=\left|\int_{-\infty}^{0}\textrm{d}t\;\mathcal{E}_{\textrm{truncated}}^{\ast}\left(t\right)\mathcal{E}_{\textrm{ideal}}\left(t\right)\right|^{2}=A^{-2}$ between the truncated pulse and its ideal version is better than $99\%$.

We found the following numerical procedure to be efficient. For each optimization run, we first set $\delta_{\textnormal{s}}(t)=\delta_{\textnormal{k}}(t)=0$
and optimize solely with respect to $\tilde{\Omega}(t)$. The step size $\lambda_{\tilde{\Omega}}$ is taken within the range of $[10^{-2},10^{2}]$ with smaller $B$ corresponding to smaller values of $\lambda_{\tilde{\Omega}}$. The initial guess for $\tilde{\Omega}(t)$ in the run with the maximal value of $J=J_{\text{max}}$ for each $B$ is a constant (square) pulse for $t\leq T'$. For smaller values of $J<J_{\text{max}}$, we attempt two independent optimization procedures using different initial guesses based on the optimal solution $\tilde{\Omega}_\mathrm{iter,1}(t)$ previously computed for the same $B$ and nearest $J$. One guess is $\tilde{\Omega}_\mathrm{iter,1}(t)$ up to the time at which $\mathcal{K}$ is maximal with additional padding of zeros at the end of the pulse to account for the increase of $T'$ due to the decrease of $J$. The other guess is a square pulse taking the average value of $\tilde{\Omega}_\mathrm{iter,1}(t)$ over the time range $t_{\ast}/2\leq t\leq 0$, where the simulation starts at time $t_{\ast}<0$. The latter guess is driven by the fact that the efficiency is not very sensitive to the early shape of the control pulse, where the exponential input is at its tail, containing only a negligible fraction of the photonic excitation. For each value of $J$ and $B$, we record the new initial guess $\tilde{\Omega}_\mathrm{iter,2}(t)$ as the solution that realizes higher storage efficiency of the two optimization attempts.
It is worth mentioning at this point that $\gamma_{\Omega}\gg J$ also functions as a term decoupling the alkali and nuclear spins, and can thus fulfill the role of $\delta_{k}$ in this first optimization step. We then, in the spin-exchange case only, allow the solver to vary $\delta_{\textnormal{s}}(t)$ and $\delta_{\textnormal{k}}(t)$
as well as $\tilde{\Omega}(t)$. We choose the steps adaptively: $\lambda^{\left(n\right)}_{\delta_{\textnormal{s}}}=\text{mean}(\gamma_{\Omega})^{-1}$
and $\lambda^{\left(n\right)}_{\delta_{\textnormal{k}}}=\text{mean}(\gamma_{J})^{-1}$, according to the values of $\gamma_{\Omega}(t)$ and $\gamma_{J}(t)$ corresponding to $\tilde{\Omega}^{\left(n-1\right)}(t)$ in the previous optimization iteration. Here we use the definitions $\gamma_{\Omega}\equiv\text{re}(\Gamma_{\Omega})$,  $\gamma_{J}\equiv\text{re}(\Gamma_{J})$,  and $\Gamma_{J}\left(t\right)\equiv{J^{2}}/({\Gamma_{\Omega}\left(t\right)+\gamma_{\text{s}}+i\delta_{\text{s}}\left(t\right)})$.

We typically observe convergence in $|\mathcal{K}(T')|^{2}$ or $|\mathcal{K}_r(T')|^{2}$ after $\sim5000$ iterations at $J=J_{\textrm{max}}$, whereas, for $J<J_{\textrm{max}}$, it is generally faster because the initial condition is already close to the optimum. We also try different initial conditions, including high values of $\delta_{\textnormal{k}}$ for the duration of the input field. We find that the shapes of $\delta_{\textnormal{k}}$ and $\delta_{\textnormal{s}}$ have only a minor influence on the obtained efficiency, except for the time the control field $\Omega(t)$ is turned off and complete decoupling between the spin ensemble is necessary, \textit{e.g.}, following the end of the storage. In such cases, a large $\delta_{\textnormal{k}}$ enables effective suppression of the coupling to the electronic spins.

\section{Storage Efficiency}
\label{appendix:structure_of_efficiency}
In this Appendix, we cast the equations of motion in normalized units, to highlight the dependence of the storage efficiency on the system parameters. We begin by identifying that 
\begin{equation}
    Q\Omega^{\ast}=\sqrt{\dfrac{\Gamma_{\Omega}}{\Gamma_{\Omega}^{\ast}}}\sqrt{\dfrac{\Omega^{\ast}}{\Omega}}\sqrt{2\gamma_{\Omega}}\sqrt{\frac{C}{C+1}},
    \label{eq:phase_of_Omega_terms}
\end{equation}
\begin{figure*}
\begin{centering}
\makebox[\textwidth]{\includegraphics[clip,scale=0.6]{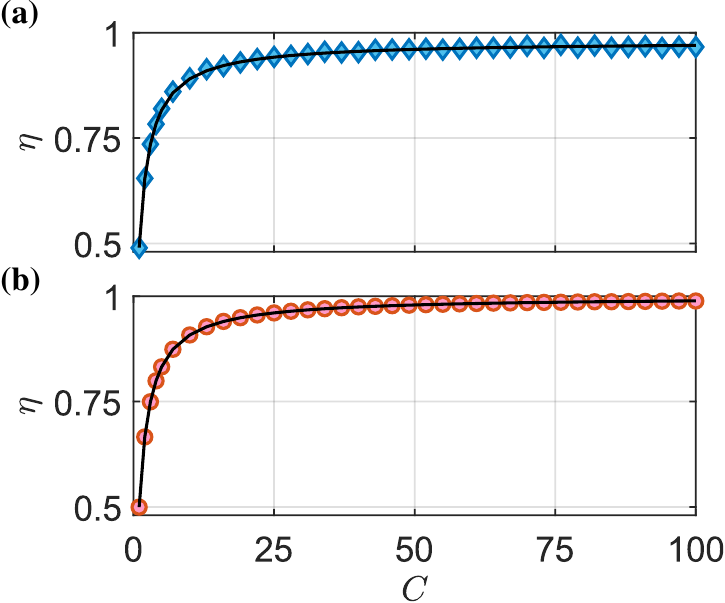}}
\par\end{centering}
\centering{}\caption{Numerical verification of the scaling with $C$. The factorization $\eta=\eta_{\infty}C/(C+1)$ (black line) is verified numerically (light blue diamond and light red circle) for two configurations: \textbf{(a)} $B/\gamma_{\textnormal{s}}=10^{3}$ [corresponding to the blue diamond in Fig.~\ref{fig:optimal_storage_strategies}(a)] and \textbf{(b)} $B/\gamma_{\textnormal{s}}\approx 5.62\cdot 10^{-2}$ [corresponding to the red circle in Fig.~\ref{fig:optimal_storage_strategies}(a)], both taken at $J/\gamma_{\textnormal{s}}=100$.
\label{fig:cooperativity}}
\end{figure*}where $\Gamma_{\Omega}$ and $Q$ were defined in the beginning of the App.~\ref{appendix:optimal_control_protocol_details}. The first two factors on the right-hand-side of this equation are phases. We therefore use the notation $e^{i\Phi_{\Omega}}=\sqrt{\Gamma_{\Omega}/\Gamma_{\Omega}^{\ast}}\sqrt{\Omega^{\ast}/\Omega}$ and take  $\Phi=\Phi_{\Omega}+\Phi_{\mathcal{E}}$ as the total phase that includes the phase of the signal, which can be cast as $\mathcal{E}_{\textrm{in}}=e^{i\Phi_{\mathcal{E}}}\left|\mathcal{E}_{\textrm{in}}\right|$. Using this notation, Eqs.\ (\ref{eq:numerical_equations_alkali},\ref{eq:numerical_equations_noble}) take the form
\begin{align}
\partial_{t}\mathcal{S} &=-\left(\gamma_{\text{s}}+\gamma_{\Omega}+i(\delta_{\text{s}}+\delta_{\Omega})\right)\mathcal{S}-iJ\mathcal{K}-e^{i\Phi}\sqrt{2\gamma_{\Omega}}\sqrt{\tfrac{C}{C+1}}\left|\mathcal{E}_{\text{in}}\right|,\label{eq:numerical_equations_alkali_with_phase}\\
\partial_{t}\mathcal{K} &=-(\gamma_{\textnormal{k}}+i\delta_{\text{k}})\mathcal{K}-i\xi J\mathcal{S},\label{eq:numerical_equations_noble_with_phase}
\end{align}
where $\delta_{\Omega}=\textrm{im}(\Gamma_{\Omega})$. To eliminate the phase factor from these equations, we switch to a rotating frame by introducing the transformation $\mathcal{S}=e^{i\Phi}\mathcal{\bar{S}}$ and $\mathcal{K}=e^{i\Phi}\mathcal{\bar{K}}$. Applying this transformation to Eqs.~(\ref{eq:numerical_equations_alkali_with_phase},\ref{eq:numerical_equations_noble_with_phase}) yields 
\begin{align}
\partial_{t}\mathcal{\bar{S}} &=-\left(\gamma_{\text{s}}+\gamma_{\Omega}+i\bar{\delta}_{\text{s}}\right)\mathcal{\bar{S}}-iJ\mathcal{\bar{K}}-\sqrt{2\gamma_{\Omega}}\sqrt{\tfrac{C}{C+1}}\left|\mathcal{E}_{\text{in}}\right|,\label{eq:numerical_equations_alkali_rotating_frame}\\
\partial_{t}\mathcal{\bar{K}} &=-\left(\gamma_{\textnormal{k}}+i\bar{\delta}_{\text{k}}\right)\mathcal{\bar{K}}-i\xi J\mathcal{\bar{S}},\label{eq:numerical_equations_noble_rotating_frame}
\end{align}
where the detunings are shifted as  $\bar{\delta}_{\textrm{s}}=\delta_{\textrm{s}}+\delta_{\Omega}+\partial_{t}\Phi$ and $\bar{\delta}_{\textrm{k}}=\delta_{\textrm{k}}+\partial_{t}\Phi$. Lastly, we make the equations dimensionless by working in units of $\gamma_{\textrm{s}}$, via $\partial_{\tau}=\gamma_{\textrm{s}}\partial_{t}$:
\begin{align}
\partial_{\tau}\mathcal{\bar{S}} &=-\left(1+\gamma_{\Omega}/\gamma_{\textrm{s}}+i\bar{\delta}_{\text{s}}/\gamma_{\textrm{s}}\right)\mathcal{\bar{S}}-i(J/\gamma_{\textrm{s}})\mathcal{\bar{K}}-\sqrt{2\gamma_{\Omega}/\gamma_{\textrm{s}}}\sqrt{\eta_{C}}\left|\mathcal{E}_{\text{in}}\right|/\sqrt{\gamma_{\textrm{s}}},\label{eq:numerical_equations_alkali_dimensionless}\\
\partial_{\tau}\mathcal{\bar{K}} &=-\left(\gamma_{\textnormal{k}}/\gamma_{\textrm{s}}+i\bar{\delta}_{\text{k}}/\gamma_{\textrm{s}}\right)\mathcal{\bar{K}}-i\xi (J/\gamma_{\textrm{s}})\mathcal{\bar{S}},\label{eq:numerical_equations_noble_dimensionless}
\end{align}
The new field $\sqrt{\eta_{C}}\left|\mathcal{E}_{\text{in}}\left(\tau\right)\right|/\sqrt{\gamma_{\textrm{s}}}$ now contains $\eta_{C}=\frac{C}{C+1}$ excitations and its bandwidth is $2B/\gamma_{\textrm{s}}$. From Eqs.\ (\ref{eq:numerical_equations_alkali_dimensionless},\ref{eq:numerical_equations_noble_dimensionless}), we can immediately infer that the efficiency assumes the form $\eta\left(C,\gamma_{\textrm{k}},\gamma_{\textrm{s}},J,B\right)=\tfrac{C}{C+1}\eta_{\infty}\left(\gamma_{\textrm{k}}/\gamma_{\textrm{s}},J/\gamma_{\textrm{s}},B/\gamma_{\textrm{s}},\xi\right)$, as used in the main text. In this work, the parameter space is spanned by two independent unitless parameters $J/\gamma_{\textrm{s}}$ and $B/\gamma_{\textrm{s}}$. For the spin-exchange configuration, we assume $\gamma_{\textrm{k}}=0$ and $\xi=1$, while for the metastability-exchange configuration, we have $\xi=-1$ and $\gamma_{\textrm{k}}/\gamma_{\textrm{s}}=r$, where the value of $r$ is determined by the parameter $J/\gamma_{\textrm{s}}=\sqrt{r}$.

We numerically verify the dependence of the numerically-optimized efficiency on the cooperativity for two distinctive different points of the spin-exchange configuration. In Fig.~\ref{fig:cooperativity}, we present the numerically attained efficiency for each value of $C$ compared with the scaled efficiency calculated at $C=100$, showing excellent agreement.

\section{Robustness to different temporal signals}
\label{appendix:dependence_on_pulse_shape}
\begin{figure*}[t!]
\begin{centering}
\makebox[\textwidth]{\includegraphics[clip,scale=0.52]{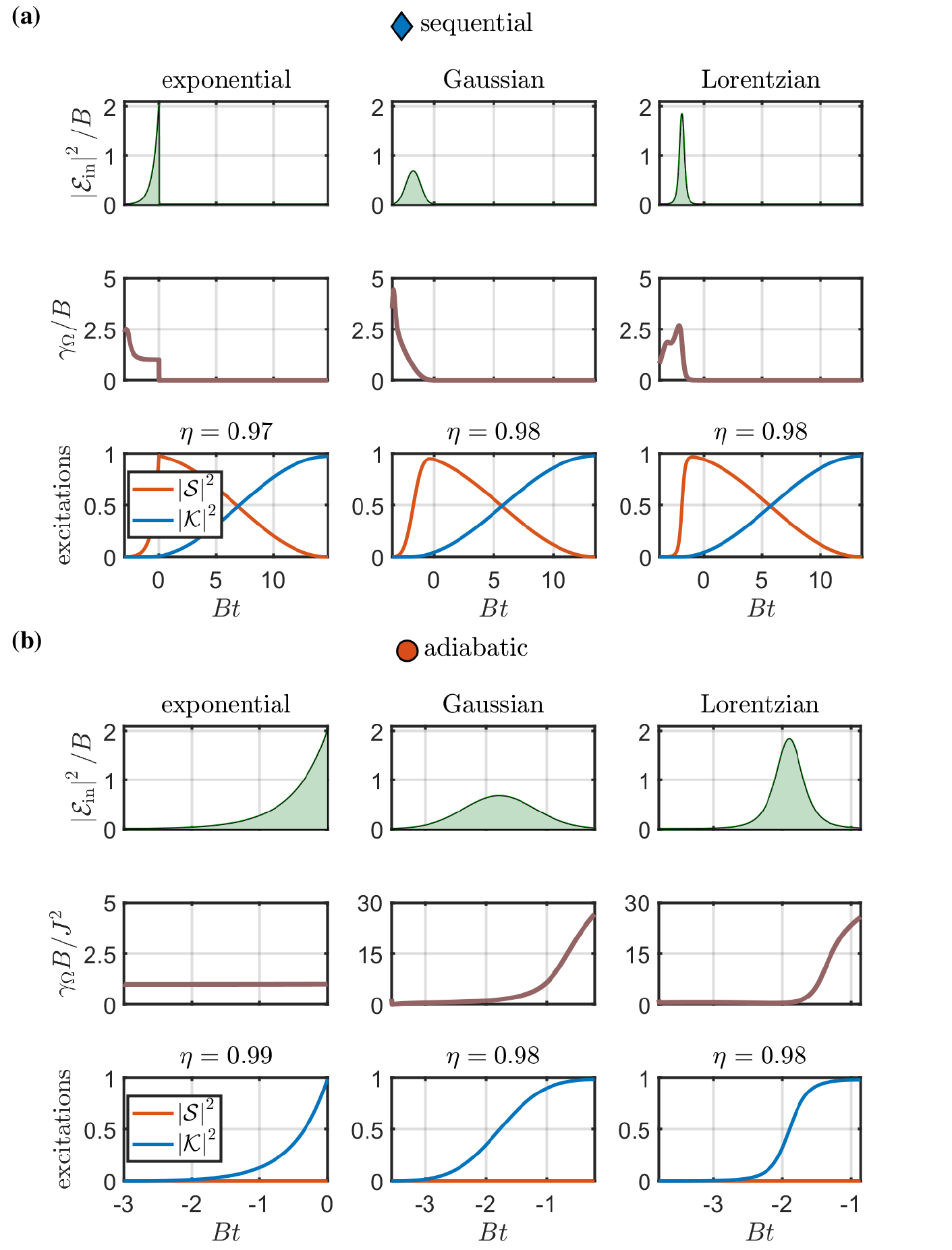}}
\par\end{centering}
\centering{}\caption{Numerical optimization for storage of exponential, Gaussian, and Lorentzian pulse shapes with bandwidth $2B$ for the spin-exchange configuration ($\xi=1$). \textbf{(a)} $J/\gamma_{\textrm{s}}\approx 100$ and $B/\gamma_{\textrm{s}}\approx 10^{3}$, corresponding to the blue diamond in Fig.~\ref{fig:optimal_storage_strategies}(a). \textbf{(b)} $J/\gamma_{\textrm{s}}\approx 100$ and $B/\gamma_{\textrm{s}}\approx 5.62\cdot 10^{-2}$, corresponding to the red circle in Fig.~\ref{fig:optimal_storage_strategies}(a). Top: optical signal. Middle:  optical control field (a parameter proportional to its intensity). Bottom:  
The efficiencies differ between different pulse shapes by no more than $1\%$.\label{fig:different_pulses}}
\end{figure*}
In this appendix, we verify,  for $\xi=1$, the robustness and applicability of our results to different pulse shapes. For that purpose, we have numerically tested optimal storage of light for different pulse shapes at several different system parameters. In Fig.~\ref{fig:different_pulses}, we present the control fields and efficiencies for two cases corresponding to the parameters of the red circle and blue diamond in Fig.~\ref{fig:optimal_storage_strategies}(a). We compare the exponential pulse with Gaussian and Lorentzian pulses of similar bandwidth, given by the following expressions:
\begin{align}
\mathcal{E}_{{\rm in}}\left(t\right) &=A\frac{1}{\left(2\pi T^{2}\right)^{\frac{1}{4}}}\exp\left(-\frac{\left(t-t_{\ast}/2\right)^{2}}{4T^{2}}\right),\,\,\,\,\,\,\,\:\,\,T=\frac{\sqrt{2\ln 2}}{2}B^{-1},
\label{eq:gaussian shaped pulse}
\\
\mathcal{E}_{{\rm in}}\left(t\right) &=A\sqrt{\frac{2}{\pi T}}\frac{T^{2}}{T^{2}+\left(t-t_{\ast}/2\right)^{2}},\,\,\,\,\,\,\,\:\,\,T=\frac{\ln 2}{2}B^{-1},
\label{eq:lorentzian shaped pulse}
\end{align}
where $t_{\star}\leq t\leq 0$ and $A$ is a normalization constant that depends on the specific pulse shape (the value in the main text is for an exponentially-shaped pulse). Both pulses are truncated symmetrically [see top row in Fig.~\ref{fig:different_pulses}(a) and Fig.~\ref{fig:different_pulses}(b)], and we choose $t_*$ such that the pulses have a $99\%$ overlap with the ideal (non-truncated) pulse shape, similar to the exponential pulse shape presented in Appendix \ref{appendix:optimal_control_protocol_details}. For this test, we have optimized over $\gamma_{\Omega}$, without the additional step of optimizing over $\delta_{\rm s}$ and $\delta_{\rm k}$ as done for the exponential pulse, setting the latter to zero. This step can be justified by the observation that $\gamma_{\Omega}>J$ decouples alkali and nuclear spins similarly to $\delta_{k}$, as apparent in all numerical solutions in this appendix. We furthermore truncate the solution at time $T^{\prime}$ where the storage efficiency is maximal (resulting, \textit{e.g.}, in the asymmetry of the Lorentzian curve in Fig.\ \ref{fig:different_pulses}(b)). Except for different temporal shaping of the control fields, we find similar performance for different pulses. The predicated values for the efficiencies, calculated using Eqs.~(\ref{eq:memory efficiency adiabatic}-\ref{eq:efficiency of the fast storage scheme}), $\eta_{\textrm{sequential}}=97\%$ and $\eta_{\textrm{adiabatic}}=99\%$,  respectively, are within $1\%$ of the values obtained from optimization.

\section{\label{sec:optimized_delta_S} Optimal $\delta_{\textnormal{s}}$}

In this appendix, we analyze the dependence of the sequential and adiabatic storage schemes on $\delta_{\textnormal{s}}$ for the spin-exchange configuration, to highlight its role in the storage process. We show analytically that $\delta_\textrm{s}=0$ is optimal for both the sequential and adiabatic schemes, consistent with our numerical finding that $\delta_{\textrm{s}}=0$ remains optimal across the explored parameter space. In the sequential scheme, under the assumption that the noble-gas spins are decoupled ($\left|\delta_{\textrm{k}}-\delta_{\textrm{s}}\right|\gg J$) in the first stage of the storage (${\mathcal{E}}_{\text{in}}\rightarrow{\mathcal{S}}$), the spin coherence is given by \cite{Gorshkov1,noble-gas-PRA} 
\begin{equation}\mathcal{S}\left(0\right)=\displaystyle\int_{-\infty}^{0}h_{\Omega}\left(t\right)\mathcal{E}_{\textrm{in}}\left(t\right)\textrm{d}t.
\end{equation}
\noindent Here, the transfer function is $h_{\Omega}\left(t\right)=-Q\Omega^{\ast}e^{-\int_{t}^{0}\left[\gamma_{\textrm{s}}+\Gamma_{\Omega}\left(s\right)+i\delta_{\textrm{s}}\right]\textrm{d}s}$ with $Q$ and $\Gamma_{\Omega}$ defined in App.~\ref{appendix:optimal_control_protocol_details}. Since the efficiency has the form of an inner product between the function $h_{\Omega}$ and $\mathcal{E}_{\text{in}}^{*}\left(t\right)$,
the maximal overlap appears for $h_{\Omega}(t)\propto\mathcal{E}_{\text{in}}^{*}\left(t\right)$
\citep{Gorshkov1}. When both $\mathcal{E}_{\text{in}}\left(t\right)$
{[}Eq.~(\ref{eq:exponentially shaped pulse}){]} and $\Omega$ are
real functions and $\Delta=0$, we obtain the condition
\begin{equation}
\text{im}(h_{\Omega}(t))=0,
\end{equation}
which is satisfied at all time for $\delta_{\textnormal{s}}=0$. Note that, in principle, non-zero $\delta_{\textrm{s}}$ can counter non-zero $\Delta$ by setting at all times $\operatorname{im}(\Gamma_{\Omega})+\delta_{\textrm{s}}=0$, because then $\Gamma_{\Omega}+i\delta_{\textrm{s}}=\gamma_{\Omega}+i\left[\operatorname{im}(\Gamma_{\Omega})+\delta_{\textrm{s}}\right]=\gamma_{\Omega}$ is real. However, since $\gamma_{\Omega}\leq|\Gamma_{\Omega}|$ (with equality if-and-only-if $\Delta=0$), one has to increase $\Omega$ to compensate for this reduction. It is therefore always preferable to set $\Delta=0$, maximizing the dynamic range of $\gamma_{\Omega}$. This is especially important if $|\Omega(t)|\leq\Omega_{\textrm{max}}$ is bounded (due to experimental constraints, for example).

In
the second stage of the storage (${\mathcal{S}}\rightarrow{\mathcal{K}}$),
the exchange evolution depends on the exchange rate $\tilde{J}(\delta)=\sqrt{J^{2}+\left(\delta+i\gamma_{\textrm{s}}\right)^{2}/4}$, which in turn depends
only on $\delta=\delta_{\textrm{k}}-\delta_{\textrm{s}}$ \cite{noble-gas-PRA} and
is optimal for $\delta=0$. Therefore, fixing $\delta_{\textnormal{s}}=\delta_{\textnormal{k}}=0$
attains that optimum. We therefore conclude that the choice of $\delta_{\textnormal{s}}=0$
maximizes the efficiency of sequential storage.

\begin{figure*}[t]
\begin{centering}
\includegraphics[clip,scale=0.55]{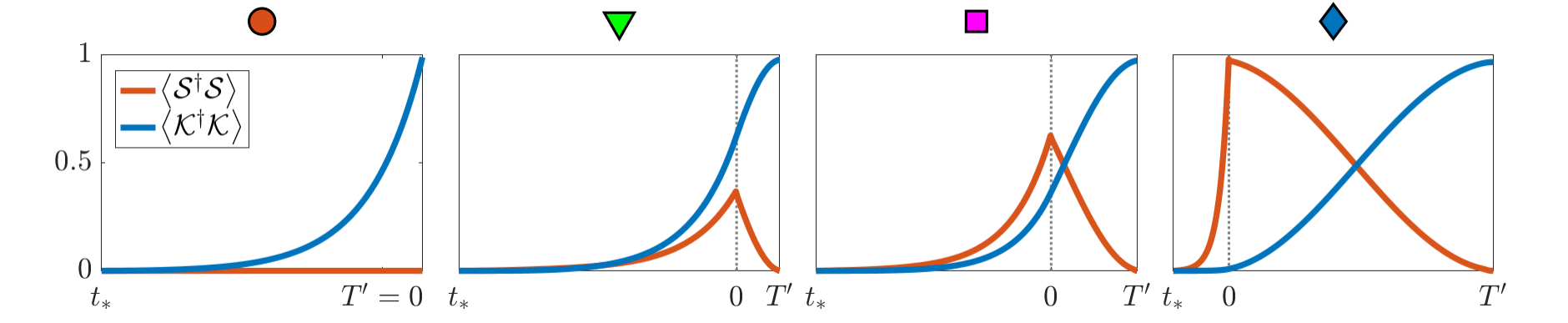}\par\end{centering}
\centering{}\caption{\textbf{Intermediate solutions in spin-exchange-based memories}. 
{We show the dynamics of the alkali-metal  $\langle\hat{\mathcal{S}}^{\dagger}\hat{\mathcal{S}}\rangle$ (red) and noble-gas  $\langle\hat{\mathcal{K}}^{\dagger}\mathcal{K}\rangle$ (blue) collective spin excitations  in the strong coupling-regime, for the spin-exchange based configuration. The spins are driven by the numerically-optimized control fields for an exponentially-shaped input pulse, as one traces a line in parameter space, from adiabatic to sequential regime. The different graphs correspond to the markers in Fig.\ \ref{fig:optimal_storage_strategies}(d).}
\label{fig:SE_intermediate_solutions}}
\end{figure*}
In the adiabatic scheme, after adiabatic elimination of the electron spin coherence in the low-bandwidth limit, the nuclear spin coherence is given by \cite{noble-gas-PRA}
\begin{equation}  \mathcal{K}\left(0\right)=\displaystyle\int_{-\infty}^{0}h_{J}\left(t\right)\mathcal{E}_{\textrm{in}}\left(t\right)\textrm{d}t,
\end{equation}
\noindent with $h_{J}\left(t\right)=\xi a_{J}e^{-\int_{t}^{0}\left[\gamma_{\textrm{k}}+\xi\Gamma_{J}\left(s\right)+i\delta_{\textrm{k}}\right]\textrm{d}s}$, where $\Gamma_{J}=J^{2}/[\Gamma_{\Omega}+\gamma_{\rm s}+i\delta_{\rm s}]$ and $a_{J}=iQ\Omega^{\ast}\Gamma_{J}/J$.
Maximal storage efficiency is obtained when
this integral is maximized. Considering
this integral as an inner product, the maximum is now attained for
$h_{J}(t)\propto\mathcal{E}_{\text{in}}^{*}\left(t\right)$. For real $\mathcal{E}_{\text{in}}\left(t\right),\,\Omega$
and for $\Delta=0$, the condition is now
\begin{equation}
\text{im}(h_{J}(t))=0.
\end{equation}
Here $\delta_{\textnormal{s}}$ takes the role of $\Delta$ in a standard $\Lambda$-system storage \cite{Gorshkov1}. However, in contrast to a standard $\Lambda$-system, $J$ is a constant making $\text{re}(\Gamma_{J})\leq\frac{J^{2}}{\gamma_{\Omega}+\gamma_{\textnormal{s}}}$ bounded. To exploit its full range, one must set $\delta_{\textnormal{s}}=0$, saturating the inequality. Consequently $\Gamma_{J}$ is real (and $a_{J}$ has a \textbf{constant} phase), making the choice $\delta_{\textnormal{k}}=0$ optimal. The results in this appendix are consistent with our numerical results.

\section{Intermediate solutions in spin-exchange-based memories}
\label{appendix:SE_intermediate_solutions}
{It is also interesting to consider the numerical solution when crossing from the ultra-low bandwidth to high bandwidth, when the optimal solutions change from the adiabatic regime to the sequential regime, respectively. In this appendix, we specifically focus on spin-exchange-based memories in the strong-coupling regime ($J\gg\gamma_{\textrm{s}}$) because they can maintain high storage efficiency for such variable bandwidth. In Fig.~\ref{fig:SE_intermediate_solutions},  we show the temporal shape of the electronic $\langle\hat{\mathcal{S}}^{\dagger}\hat{\mathcal{S}}\rangle$ and nuclear $\langle\hat{\mathcal{K}}^{\dagger}\mathcal{K}\rangle$ collective spin excitations for the different pulse bandwidths $B$ corresponding to the symbols in Fig.~\ref{fig:optimal_storage_strategies}(d). The leftmost plot shows a purely adiabatic solution where the low-bandwidth input field is slowly stored directly onto the noble-gas spins. Conversely, the rightmost plot shows the dynamics in the sequential regime, where the high-bandwidth input field is stored first on the alkali-spin, and only later transferred onto the noble-gas spins via a resonant spin-exchange process. These two solutions were already shown in Fig.~\ref{fig:optimal_storage_strategies}(e-f). The two middle plots, on the other hand, show intermediate, mixed solutions that are neither purely adiabatic nor purely sequential. In this intermediate regime, the input field is stored first on a state that has both $\hat{\mathcal{S}}$ and $\hat{\mathcal{K}}$ characters. Later on, this state is rotated fully into $\hat{\mathcal{K}}$ via the spin-exchange coupling. Our numerical optimization shows therefore that efficient solutions can be engineered for intermediate bandwidth.

\bibliography{Refs}

\begin{thebibliography}{41}%
\makeatletter
\providecommand \@ifxundefined [1]{%
 \@ifx{#1\undefined}
}%
\providecommand \@ifnum [1]{%
 \ifnum #1\expandafter \@firstoftwo
 \else \expandafter \@secondoftwo
 \fi
}%
\providecommand \@ifx [1]{%
 \ifx #1\expandafter \@firstoftwo
 \else \expandafter \@secondoftwo
 \fi
}%
\providecommand \natexlab [1]{#1}%
\providecommand \enquote  [1]{``#1''}%
\providecommand \bibnamefont  [1]{#1}%
\providecommand \bibfnamefont [1]{#1}%
\providecommand \citenamefont [1]{#1}%
\providecommand \href@noop [0]{\@secondoftwo}%
\providecommand \href [0]{\begingroup \@sanitize@url \@href}%
\providecommand \@href[1]{\@@startlink{#1}\@@href}%
\providecommand \@@href[1]{\endgroup#1\@@endlink}%
\providecommand \@sanitize@url [0]{\catcode `\\12\catcode `\$12\catcode
  `\&12\catcode `\#12\catcode `\^12\catcode `\_12\catcode `\%12\relax}%
\providecommand \@@startlink[1]{}%
\providecommand \@@endlink[0]{}%
\providecommand \url  [0]{\begingroup\@sanitize@url \@url }%
\providecommand \@url [1]{\endgroup\@href {#1}{\urlprefix }}%
\providecommand \urlprefix  [0]{URL }%
\providecommand \Eprint [0]{\href }%
\providecommand \doibase [0]{http://dx.doi.org/}%
\providecommand \selectlanguage [0]{\@gobble}%
\providecommand \bibinfo  [0]{\@secondoftwo}%
\providecommand \bibfield  [0]{\@secondoftwo}%
\providecommand \translation [1]{[#1]}%
\providecommand \BibitemOpen [0]{}%
\providecommand \bibitemStop [0]{}%
\providecommand \bibitemNoStop [0]{.\EOS\space}%
\providecommand \EOS [0]{\spacefactor3000\relax}%
\providecommand \BibitemShut  [1]{\csname bibitem#1\endcsname}%
\let\auto@bib@innerbib\@empty
%</preamble>
\bibitem [{\citenamefont {Lvovsky}\ \emph {et~al.}(2009)\citenamefont
  {Lvovsky}, \citenamefont {Sanders},\ and\ \citenamefont
  {Tittel}}]{Tittel-review-nature-photonics}%
  \BibitemOpen
  \bibfield  {author} {\bibinfo {author} {\bibfnamefont {A.~I.}\ \bibnamefont
  {Lvovsky}}, \bibinfo {author} {\bibfnamefont {B.~C.}\ \bibnamefont
  {Sanders}}, \ and\ \bibinfo {author} {\bibfnamefont {W.}~\bibnamefont
  {Tittel}},\ }\href@noop {} {\bibfield  {journal} {\bibinfo  {journal} {Nature
  Photonics}\ }\textbf {\bibinfo {volume} {3}},\ \bibinfo {pages} {706}
  (\bibinfo {year} {2009})}\BibitemShut {NoStop}%
\bibitem [{\citenamefont {O'brien}(2007)}]{oPTICAL-QUANTUM-COMPUTING-SCIENCE}%
  \BibitemOpen
  \bibfield  {author} {\bibinfo {author} {\bibfnamefont {J.~L.}\ \bibnamefont
  {O'brien}},\ }\href@noop {} {\bibfield  {journal} {\bibinfo  {journal}
  {Science}\ }\textbf {\bibinfo {volume} {318}},\ \bibinfo {pages} {1567}
  (\bibinfo {year} {2007})}\BibitemShut {NoStop}%
\bibitem [{\citenamefont {Heshami}\ \emph {et~al.}(2016)\citenamefont
  {Heshami}, \citenamefont {England}, \citenamefont {Humphreys}, \citenamefont
  {Bustard}, \citenamefont {Acosta}, \citenamefont {Nunn},\ and\ \citenamefont
  {Sussman}}]{Heshami-2016}%
  \BibitemOpen
  \bibfield  {author} {\bibinfo {author} {\bibfnamefont {K.}~\bibnamefont
  {Heshami}}, \bibinfo {author} {\bibfnamefont {D.~G.}\ \bibnamefont
  {England}}, \bibinfo {author} {\bibfnamefont {P.~C.}\ \bibnamefont
  {Humphreys}}, \bibinfo {author} {\bibfnamefont {P.~J.}\ \bibnamefont
  {Bustard}}, \bibinfo {author} {\bibfnamefont {V.~M.}\ \bibnamefont {Acosta}},
  \bibinfo {author} {\bibfnamefont {J.}~\bibnamefont {Nunn}}, \ and\ \bibinfo
  {author} {\bibfnamefont {B.~J.}\ \bibnamefont {Sussman}},\ }\href@noop {}
  {\bibfield  {journal} {\bibinfo  {journal} {Journal of Modern Optics}\
  }\textbf {\bibinfo {volume} {63}},\ \bibinfo {pages} {2005} (\bibinfo {year}
  {2016})}\BibitemShut {NoStop}%
\bibitem [{\citenamefont {Sangouard}\ \emph {et~al.}(2011)\citenamefont
  {Sangouard}, \citenamefont {Simon}, \citenamefont {De~Riedmatten},\ and\
  \citenamefont {Gisin}}]{Gisin-repeaters}%
  \BibitemOpen
  \bibfield  {author} {\bibinfo {author} {\bibfnamefont {N.}~\bibnamefont
  {Sangouard}}, \bibinfo {author} {\bibfnamefont {C.}~\bibnamefont {Simon}},
  \bibinfo {author} {\bibfnamefont {H.}~\bibnamefont {De~Riedmatten}}, \ and\
  \bibinfo {author} {\bibfnamefont {N.}~\bibnamefont {Gisin}},\ }\href@noop {}
  {\bibfield  {journal} {\bibinfo  {journal} {Reviews of Modern Physics}\
  }\textbf {\bibinfo {volume} {83}},\ \bibinfo {pages} {33} (\bibinfo {year}
  {2011})}\BibitemShut {NoStop}%
\bibitem [{\citenamefont {Hammerer}\ \emph {et~al.}(2010)\citenamefont
  {Hammerer}, \citenamefont {S{\o}rensen},\ and\ \citenamefont
  {Polzik}}]{Polzik-RMP-2010}%
  \BibitemOpen
  \bibfield  {author} {\bibinfo {author} {\bibfnamefont {K.}~\bibnamefont
  {Hammerer}}, \bibinfo {author} {\bibfnamefont {A.~S.}\ \bibnamefont
  {S{\o}rensen}}, \ and\ \bibinfo {author} {\bibfnamefont {E.~S.}\ \bibnamefont
  {Polzik}},\ }\href@noop {} {\bibfield  {journal} {\bibinfo  {journal}
  {Reviews of Modern Physics}\ }\textbf {\bibinfo {volume} {82}},\ \bibinfo
  {pages} {1041} (\bibinfo {year} {2010})}\BibitemShut {NoStop}%
\bibitem [{\citenamefont {Gemmel}\ \emph {et~al.}(2010)\citenamefont {Gemmel},
  \citenamefont {Heil}, \citenamefont {Karpuk}, \citenamefont {Lenz},
  \citenamefont {Ludwig}, \citenamefont {Sobolev}, \citenamefont {Tullney},
  \citenamefont {Burghoff}, \citenamefont {Kilian}, \citenamefont
  {Knappe-Gr{\"u}neberg} \emph
  {et~al.}}]{Gemmel-60-hours-coherence-time-He-2010}%
  \BibitemOpen
  \bibfield  {author} {\bibinfo {author} {\bibfnamefont {C.}~\bibnamefont
  {Gemmel}}, \bibinfo {author} {\bibfnamefont {W.}~\bibnamefont {Heil}},
  \bibinfo {author} {\bibfnamefont {S.}~\bibnamefont {Karpuk}}, \bibinfo
  {author} {\bibfnamefont {K.}~\bibnamefont {Lenz}}, \bibinfo {author}
  {\bibfnamefont {C.}~\bibnamefont {Ludwig}}, \bibinfo {author} {\bibfnamefont
  {Y.}~\bibnamefont {Sobolev}}, \bibinfo {author} {\bibfnamefont
  {K.}~\bibnamefont {Tullney}}, \bibinfo {author} {\bibfnamefont
  {M.}~\bibnamefont {Burghoff}}, \bibinfo {author} {\bibfnamefont
  {W.}~\bibnamefont {Kilian}}, \bibinfo {author} {\bibfnamefont
  {S.}~\bibnamefont {Knappe-Gr{\"u}neberg}},  \emph {et~al.},\ }\href@noop {}
  {\bibfield  {journal} {\bibinfo  {journal} {The European Physical Journal D}\
  }\textbf {\bibinfo {volume} {57}},\ \bibinfo {pages} {303} (\bibinfo {year}
  {2010})}\BibitemShut {NoStop}%
\bibitem [{\citenamefont {Gentile}\ \emph
  {et~al.}(2017{\natexlab{a}})\citenamefont {Gentile}, \citenamefont {Nacher},
  \citenamefont {Saam},\ and\ \citenamefont {Walker}}]{Walker-RMP-2017}%
  \BibitemOpen
  \bibfield  {author} {\bibinfo {author} {\bibfnamefont {T.~R.}\ \bibnamefont
  {Gentile}}, \bibinfo {author} {\bibfnamefont {P.}~\bibnamefont {Nacher}},
  \bibinfo {author} {\bibfnamefont {B.}~\bibnamefont {Saam}}, \ and\ \bibinfo
  {author} {\bibfnamefont {T.}~\bibnamefont {Walker}},\ }\href@noop {}
  {\bibfield  {journal} {\bibinfo  {journal} {Reviews of Modern Physics}\
  }\textbf {\bibinfo {volume} {89}},\ \bibinfo {pages} {045004} (\bibinfo
  {year} {2017}{\natexlab{a}})}\BibitemShut {NoStop}%
\bibitem [{\citenamefont {Katz}\ \emph {et~al.}(2020)\citenamefont {Katz},
  \citenamefont {Shaham}, \citenamefont {Polzik},\ and\ \citenamefont
  {Firstenberg}}]{Firstenberg-QND}%
  \BibitemOpen
  \bibfield  {author} {\bibinfo {author} {\bibfnamefont {O.}~\bibnamefont
  {Katz}}, \bibinfo {author} {\bibfnamefont {R.}~\bibnamefont {Shaham}},
  \bibinfo {author} {\bibfnamefont {E.~S.}\ \bibnamefont {Polzik}}, \ and\
  \bibinfo {author} {\bibfnamefont {O.}~\bibnamefont {Firstenberg}},\
  }\href@noop {} {\bibfield  {journal} {\bibinfo  {journal} {Physical Review
  Letters}\ }\textbf {\bibinfo {volume} {124}},\ \bibinfo {pages} {043602}
  (\bibinfo {year} {2020})}\BibitemShut {NoStop}%
\bibitem [{\citenamefont {Reinaudi}\ \emph {et~al.}(2007)\citenamefont
  {Reinaudi}, \citenamefont {Sinatra}, \citenamefont {Dantan},\ and\
  \citenamefont {Pinard}}]{Sinatra-squeezing}%
  \BibitemOpen
  \bibfield  {author} {\bibinfo {author} {\bibfnamefont {G.}~\bibnamefont
  {Reinaudi}}, \bibinfo {author} {\bibfnamefont {A.}~\bibnamefont {Sinatra}},
  \bibinfo {author} {\bibfnamefont {A.}~\bibnamefont {Dantan}}, \ and\ \bibinfo
  {author} {\bibfnamefont {M.}~\bibnamefont {Pinard}},\ }\href@noop {}
  {\bibfield  {journal} {\bibinfo  {journal} {Journal of Modern Optics}\
  }\textbf {\bibinfo {volume} {54}},\ \bibinfo {pages} {675} (\bibinfo {year}
  {2007})}\BibitemShut {NoStop}%
\bibitem [{\citenamefont {Serafin}\ \emph {et~al.}(2021)\citenamefont
  {Serafin}, \citenamefont {Fadel}, \citenamefont {Treutlein},\ and\
  \citenamefont {Sinatra}}]{Sinatra-squeezing2}%
  \BibitemOpen
  \bibfield  {author} {\bibinfo {author} {\bibfnamefont {A.}~\bibnamefont
  {Serafin}}, \bibinfo {author} {\bibfnamefont {M.}~\bibnamefont {Fadel}},
  \bibinfo {author} {\bibfnamefont {P.}~\bibnamefont {Treutlein}}, \ and\
  \bibinfo {author} {\bibfnamefont {A.}~\bibnamefont {Sinatra}},\ }\href@noop
  {} {\bibfield  {journal} {\bibinfo  {journal} {Physical Review Letters}\
  }\textbf {\bibinfo {volume} {127}},\ \bibinfo {pages} {013601} (\bibinfo
  {year} {2021})}\BibitemShut {NoStop}%
\bibitem [{\citenamefont {Dantan}\ \emph {et~al.}(2005)\citenamefont {Dantan},
  \citenamefont {Reinaudi}, \citenamefont {Sinatra}, \citenamefont {Lalo{\"e}},
  \citenamefont {Giacobino},\ and\ \citenamefont
  {Pinard}}]{Sinatra-2005-Metastable}%
  \BibitemOpen
  \bibfield  {author} {\bibinfo {author} {\bibfnamefont {A.}~\bibnamefont
  {Dantan}}, \bibinfo {author} {\bibfnamefont {G.}~\bibnamefont {Reinaudi}},
  \bibinfo {author} {\bibfnamefont {A.}~\bibnamefont {Sinatra}}, \bibinfo
  {author} {\bibfnamefont {F.}~\bibnamefont {Lalo{\"e}}}, \bibinfo {author}
  {\bibfnamefont {E.}~\bibnamefont {Giacobino}}, \ and\ \bibinfo {author}
  {\bibfnamefont {M.}~\bibnamefont {Pinard}},\ }\href@noop {} {\bibfield
  {journal} {\bibinfo  {journal} {Physical Review Letters}\ }\textbf {\bibinfo
  {volume} {95}},\ \bibinfo {pages} {123002} (\bibinfo {year}
  {2005})}\BibitemShut {NoStop}%
\bibitem [{\citenamefont {Walker}\ and\ \citenamefont
  {Happer}(1997)}]{SEOP-Happer-RMP}%
  \BibitemOpen
  \bibfield  {author} {\bibinfo {author} {\bibfnamefont {T.~G.}\ \bibnamefont
  {Walker}}\ and\ \bibinfo {author} {\bibfnamefont {W.}~\bibnamefont
  {Happer}},\ }\href@noop {} {\bibfield  {journal} {\bibinfo  {journal}
  {Reviews of Modern Physics}\ }\textbf {\bibinfo {volume} {69}},\ \bibinfo
  {pages} {629} (\bibinfo {year} {1997})}\BibitemShut {NoStop}%
\bibitem [{\citenamefont {Katz}\ \emph
  {et~al.}(2022{\natexlab{a}})\citenamefont {Katz}, \citenamefont {Shaham},\
  and\ \citenamefont {Firstenberg}}]{Firstenberg-Weak-collisions}%
  \BibitemOpen
  \bibfield  {author} {\bibinfo {author} {\bibfnamefont {O.}~\bibnamefont
  {Katz}}, \bibinfo {author} {\bibfnamefont {R.}~\bibnamefont {Shaham}}, \ and\
  \bibinfo {author} {\bibfnamefont {O.}~\bibnamefont {Firstenberg}},\
  }\href@noop {} {\bibfield  {journal} {\bibinfo  {journal} {PRX Quantum}\
  }\textbf {\bibinfo {volume} {3}},\ \bibinfo {pages} {010305} (\bibinfo {year}
  {2022}{\natexlab{a}})}\BibitemShut {NoStop}%
\bibitem [{\citenamefont {Katz}\ \emph
  {et~al.}(2022{\natexlab{b}})\citenamefont {Katz}, \citenamefont {Shaham},
  \citenamefont {Reches}, \citenamefont {Gorshkov},\ and\ \citenamefont
  {Firstenberg}}]{noble-gas-PRA}%
  \BibitemOpen
  \bibfield  {author} {\bibinfo {author} {\bibfnamefont {O.}~\bibnamefont
  {Katz}}, \bibinfo {author} {\bibfnamefont {R.}~\bibnamefont {Shaham}},
  \bibinfo {author} {\bibfnamefont {E.}~\bibnamefont {Reches}}, \bibinfo
  {author} {\bibfnamefont {A.~V.}\ \bibnamefont {Gorshkov}}, \ and\ \bibinfo
  {author} {\bibfnamefont {O.}~\bibnamefont {Firstenberg}},\ }\href@noop {}
  {\bibfield  {journal} {\bibinfo  {journal} {Physical Review A}\ }\textbf
  {\bibinfo {volume} {105}},\ \bibinfo {pages} {042606} (\bibinfo {year}
  {2022}{\natexlab{b}})}\BibitemShut {NoStop}%
\bibitem [{\citenamefont {Gorshkov}\ \emph
  {et~al.}(2007{\natexlab{a}})\citenamefont {Gorshkov}, \citenamefont
  {Andr{\'e}}, \citenamefont {Fleischhauer}, \citenamefont {S{\o}rensen},\ and\
  \citenamefont {Lukin}}]{Gorshkov-PRL}%
  \BibitemOpen
  \bibfield  {author} {\bibinfo {author} {\bibfnamefont {A.~V.}\ \bibnamefont
  {Gorshkov}}, \bibinfo {author} {\bibfnamefont {A.}~\bibnamefont {Andr{\'e}}},
  \bibinfo {author} {\bibfnamefont {M.}~\bibnamefont {Fleischhauer}}, \bibinfo
  {author} {\bibfnamefont {A.~S.}\ \bibnamefont {S{\o}rensen}}, \ and\ \bibinfo
  {author} {\bibfnamefont {M.~D.}\ \bibnamefont {Lukin}},\ }\href@noop {}
  {\bibfield  {journal} {\bibinfo  {journal} {Physical Review Letters}\
  }\textbf {\bibinfo {volume} {98}},\ \bibinfo {pages} {123601} (\bibinfo
  {year} {2007}{\natexlab{a}})}\BibitemShut {NoStop}%
\bibitem [{\citenamefont {Gorshkov}\ \emph
  {et~al.}(2007{\natexlab{b}})\citenamefont {Gorshkov}, \citenamefont
  {Andr{\'e}}, \citenamefont {Lukin},\ and\ \citenamefont
  {S{\o}rensen}}]{Gorshkov1}%
  \BibitemOpen
  \bibfield  {author} {\bibinfo {author} {\bibfnamefont {A.~V.}\ \bibnamefont
  {Gorshkov}}, \bibinfo {author} {\bibfnamefont {A.}~\bibnamefont {Andr{\'e}}},
  \bibinfo {author} {\bibfnamefont {M.~D.}\ \bibnamefont {Lukin}}, \ and\
  \bibinfo {author} {\bibfnamefont {A.~S.}\ \bibnamefont {S{\o}rensen}},\
  }\href@noop {} {\bibfield  {journal} {\bibinfo  {journal} {Physical Review
  A}\ }\textbf {\bibinfo {volume} {76}},\ \bibinfo {pages} {033804} (\bibinfo
  {year} {2007}{\natexlab{b}})}\BibitemShut {NoStop}%
\bibitem [{\citenamefont {Kollath-B{\"o}nig}\ \emph {et~al.}(2024)\citenamefont
  {Kollath-B{\"o}nig}, \citenamefont {Dellantonio}, \citenamefont {Giannelli},
  \citenamefont {Schmit}, \citenamefont {Morigi},\ and\ \citenamefont
  {S{\o}rensen}}]{kollath2024fast}%
  \BibitemOpen
  \bibfield  {author} {\bibinfo {author} {\bibfnamefont {J.~S.}\ \bibnamefont
  {Kollath-B{\"o}nig}}, \bibinfo {author} {\bibfnamefont {L.}~\bibnamefont
  {Dellantonio}}, \bibinfo {author} {\bibfnamefont {L.}~\bibnamefont
  {Giannelli}}, \bibinfo {author} {\bibfnamefont {T.}~\bibnamefont {Schmit}},
  \bibinfo {author} {\bibfnamefont {G.}~\bibnamefont {Morigi}}, \ and\ \bibinfo
  {author} {\bibfnamefont {A.~S.}\ \bibnamefont {S{\o}rensen}},\ }\href@noop {}
  {\bibfield  {journal} {\bibinfo  {journal} {arXiv preprint arXiv:2401.17394}\
  } (\bibinfo {year} {2024})}\BibitemShut {NoStop}%
\bibitem [{\citenamefont {Vasilev}\ \emph {et~al.}(2010)\citenamefont
  {Vasilev}, \citenamefont {Ljunggren},\ and\ \citenamefont
  {Kuhn}}]{vasilev2010single}%
  \BibitemOpen
  \bibfield  {author} {\bibinfo {author} {\bibfnamefont {G.~S.}\ \bibnamefont
  {Vasilev}}, \bibinfo {author} {\bibfnamefont {D.}~\bibnamefont {Ljunggren}},
  \ and\ \bibinfo {author} {\bibfnamefont {A.}~\bibnamefont {Kuhn}},\
  }\href@noop {} {\bibfield  {journal} {\bibinfo  {journal} {New Journal of
  Physics}\ }\textbf {\bibinfo {volume} {12}},\ \bibinfo {pages} {063024}
  (\bibinfo {year} {2010})}\BibitemShut {NoStop}%
\bibitem [{\citenamefont {Utsugi}\ \emph {et~al.}(2022)\citenamefont {Utsugi},
  \citenamefont {Goban}, \citenamefont {Tokunaga}, \citenamefont {Goto},\ and\
  \citenamefont {Aoki}}]{utsugi2022gaussian}%
  \BibitemOpen
  \bibfield  {author} {\bibinfo {author} {\bibfnamefont {T.}~\bibnamefont
  {Utsugi}}, \bibinfo {author} {\bibfnamefont {A.}~\bibnamefont {Goban}},
  \bibinfo {author} {\bibfnamefont {Y.}~\bibnamefont {Tokunaga}}, \bibinfo
  {author} {\bibfnamefont {H.}~\bibnamefont {Goto}}, \ and\ \bibinfo {author}
  {\bibfnamefont {T.}~\bibnamefont {Aoki}},\ }\href@noop {} {\bibfield
  {journal} {\bibinfo  {journal} {Physical Review A}\ }\textbf {\bibinfo
  {volume} {106}},\ \bibinfo {pages} {023712} (\bibinfo {year}
  {2022})}\BibitemShut {NoStop}%
\bibitem [{\citenamefont {Phillips}\ \emph {et~al.}(2001)\citenamefont
  {Phillips}, \citenamefont {Fleischhauer}, \citenamefont {Mair}, \citenamefont
  {Walsworth},\ and\ \citenamefont {Lukin}}]{Lukin-PRL-2001}%
  \BibitemOpen
  \bibfield  {author} {\bibinfo {author} {\bibfnamefont {D.~F.}\ \bibnamefont
  {Phillips}}, \bibinfo {author} {\bibfnamefont {A.}~\bibnamefont
  {Fleischhauer}}, \bibinfo {author} {\bibfnamefont {A.}~\bibnamefont {Mair}},
  \bibinfo {author} {\bibfnamefont {R.~L.}\ \bibnamefont {Walsworth}}, \ and\
  \bibinfo {author} {\bibfnamefont {M.~D.}\ \bibnamefont {Lukin}},\ }\href@noop
  {} {\bibfield  {journal} {\bibinfo  {journal} {Physical Review Letters}\
  }\textbf {\bibinfo {volume} {86}},\ \bibinfo {pages} {783} (\bibinfo {year}
  {2001})}\BibitemShut {NoStop}%
\bibitem [{\citenamefont {Lukin}(2003)}]{Lukin-RMP-2003}%
  \BibitemOpen
  \bibfield  {author} {\bibinfo {author} {\bibfnamefont {M.}~\bibnamefont
  {Lukin}},\ }\href@noop {} {\bibfield  {journal} {\bibinfo  {journal} {Reviews
  of Modern Physics}\ }\textbf {\bibinfo {volume} {75}},\ \bibinfo {pages}
  {457} (\bibinfo {year} {2003})}\BibitemShut {NoStop}%
\bibitem [{\citenamefont {Hosseini}\ \emph {et~al.}(2011)\citenamefont
  {Hosseini}, \citenamefont {Sparkes}, \citenamefont {Campbell}, \citenamefont
  {Lam},\ and\ \citenamefont {Buchler}}]{Buchler-GEM}%
  \BibitemOpen
  \bibfield  {author} {\bibinfo {author} {\bibfnamefont {M.}~\bibnamefont
  {Hosseini}}, \bibinfo {author} {\bibfnamefont {B.~M.}\ \bibnamefont
  {Sparkes}}, \bibinfo {author} {\bibfnamefont {G.}~\bibnamefont {Campbell}},
  \bibinfo {author} {\bibfnamefont {P.~K.}\ \bibnamefont {Lam}}, \ and\
  \bibinfo {author} {\bibfnamefont {B.~C.}\ \bibnamefont {Buchler}},\
  }\href@noop {} {\bibfield  {journal} {\bibinfo  {journal} {Nature
  Communications}\ }\textbf {\bibinfo {volume} {2}},\ \bibinfo {pages} {174}
  (\bibinfo {year} {2011})}\BibitemShut {NoStop}%
\bibitem [{\citenamefont {Julsgaard}\ \emph {et~al.}(2004)\citenamefont
  {Julsgaard}, \citenamefont {Sherson}, \citenamefont {Cirac}, \citenamefont
  {Fiur{\'a}{\v{s}}ek},\ and\ \citenamefont
  {Polzik}}]{Polzik-coherent-state-memory}%
  \BibitemOpen
  \bibfield  {author} {\bibinfo {author} {\bibfnamefont {B.}~\bibnamefont
  {Julsgaard}}, \bibinfo {author} {\bibfnamefont {J.}~\bibnamefont {Sherson}},
  \bibinfo {author} {\bibfnamefont {J.~I.}\ \bibnamefont {Cirac}}, \bibinfo
  {author} {\bibfnamefont {J.}~\bibnamefont {Fiur{\'a}{\v{s}}ek}}, \ and\
  \bibinfo {author} {\bibfnamefont {E.~S.}\ \bibnamefont {Polzik}},\
  }\href@noop {} {\bibfield  {journal} {\bibinfo  {journal} {Nature}\ }\textbf
  {\bibinfo {volume} {432}},\ \bibinfo {pages} {482} (\bibinfo {year}
  {2004})}\BibitemShut {NoStop}%
\bibitem [{\citenamefont {Jensen}\ \emph {et~al.}(2011)\citenamefont {Jensen},
  \citenamefont {Wasilewski}, \citenamefont {Krauter}, \citenamefont
  {Fernholz}, \citenamefont {Nielsen}, \citenamefont {Owari}, \citenamefont
  {Plenio}, \citenamefont {Serafini}, \citenamefont {Wolf},\ and\ \citenamefont
  {Polzik}}]{Polzik-squeezed-states-memory}%
  \BibitemOpen
  \bibfield  {author} {\bibinfo {author} {\bibfnamefont {K.}~\bibnamefont
  {Jensen}}, \bibinfo {author} {\bibfnamefont {W.}~\bibnamefont {Wasilewski}},
  \bibinfo {author} {\bibfnamefont {H.}~\bibnamefont {Krauter}}, \bibinfo
  {author} {\bibfnamefont {T.}~\bibnamefont {Fernholz}}, \bibinfo {author}
  {\bibfnamefont {B.}~\bibnamefont {Nielsen}}, \bibinfo {author} {\bibfnamefont
  {M.}~\bibnamefont {Owari}}, \bibinfo {author} {\bibfnamefont {M.~B.}\
  \bibnamefont {Plenio}}, \bibinfo {author} {\bibfnamefont {A.}~\bibnamefont
  {Serafini}}, \bibinfo {author} {\bibfnamefont {M.}~\bibnamefont {Wolf}}, \
  and\ \bibinfo {author} {\bibfnamefont {E.}~\bibnamefont {Polzik}},\
  }\href@noop {} {\bibfield  {journal} {\bibinfo  {journal} {Nature Physics}\
  }\textbf {\bibinfo {volume} {7}},\ \bibinfo {pages} {13} (\bibinfo {year}
  {2011})}\BibitemShut {NoStop}%
\bibitem [{\citenamefont {Saunders}\ \emph {et~al.}(2016)\citenamefont
  {Saunders}, \citenamefont {Munns}, \citenamefont {Champion}, \citenamefont
  {Qiu}, \citenamefont {Kaczmarek}, \citenamefont {Poem}, \citenamefont
  {Ledingham}, \citenamefont {Walmsley},\ and\ \citenamefont
  {Nunn}}]{Walmsley-cavity}%
  \BibitemOpen
  \bibfield  {author} {\bibinfo {author} {\bibfnamefont {D.}~\bibnamefont
  {Saunders}}, \bibinfo {author} {\bibfnamefont {J.}~\bibnamefont {Munns}},
  \bibinfo {author} {\bibfnamefont {T.}~\bibnamefont {Champion}}, \bibinfo
  {author} {\bibfnamefont {C.}~\bibnamefont {Qiu}}, \bibinfo {author}
  {\bibfnamefont {K.}~\bibnamefont {Kaczmarek}}, \bibinfo {author}
  {\bibfnamefont {E.}~\bibnamefont {Poem}}, \bibinfo {author} {\bibfnamefont
  {P.}~\bibnamefont {Ledingham}}, \bibinfo {author} {\bibfnamefont
  {I.}~\bibnamefont {Walmsley}}, \ and\ \bibinfo {author} {\bibfnamefont
  {J.}~\bibnamefont {Nunn}},\ }\href@noop {} {\bibfield  {journal} {\bibinfo
  {journal} {Physical Review Letters}\ }\textbf {\bibinfo {volume} {116}},\
  \bibinfo {pages} {090501} (\bibinfo {year} {2016})}\BibitemShut {NoStop}%
\bibitem [{\citenamefont {Katz}\ and\ \citenamefont
  {Firstenberg}(2018)}]{Katz-storage-of-light-2018}%
  \BibitemOpen
  \bibfield  {author} {\bibinfo {author} {\bibfnamefont {O.}~\bibnamefont
  {Katz}}\ and\ \bibinfo {author} {\bibfnamefont {O.}~\bibnamefont
  {Firstenberg}},\ }\href@noop {} {\bibfield  {journal} {\bibinfo  {journal}
  {Nature Communications}\ }\textbf {\bibinfo {volume} {9}},\ \bibinfo {pages}
  {2074} (\bibinfo {year} {2018})}\BibitemShut {NoStop}%
\bibitem [{\citenamefont {Lahad}\ and\ \citenamefont
  {Firstenberg}(2017)}]{Lahad-2017}%
  \BibitemOpen
  \bibfield  {author} {\bibinfo {author} {\bibfnamefont {O.}~\bibnamefont
  {Lahad}}\ and\ \bibinfo {author} {\bibfnamefont {O.}~\bibnamefont
  {Firstenberg}},\ }\href@noop {} {\bibfield  {journal} {\bibinfo  {journal}
  {Physical Review Letters}\ }\textbf {\bibinfo {volume} {119}},\ \bibinfo
  {pages} {113601} (\bibinfo {year} {2017})}\BibitemShut {NoStop}%
\bibitem [{\citenamefont {Gorshkov}\ \emph
  {et~al.}(2007{\natexlab{c}})\citenamefont {Gorshkov}, \citenamefont
  {Andr{\'e}}, \citenamefont {Lukin},\ and\ \citenamefont
  {S{\o}rensen}}]{Gorshkov2}%
  \BibitemOpen
  \bibfield  {author} {\bibinfo {author} {\bibfnamefont {A.~V.}\ \bibnamefont
  {Gorshkov}}, \bibinfo {author} {\bibfnamefont {A.}~\bibnamefont {Andr{\'e}}},
  \bibinfo {author} {\bibfnamefont {M.~D.}\ \bibnamefont {Lukin}}, \ and\
  \bibinfo {author} {\bibfnamefont {A.~S.}\ \bibnamefont {S{\o}rensen}},\
  }\href@noop {} {\bibfield  {journal} {\bibinfo  {journal} {Physical Review
  A}\ }\textbf {\bibinfo {volume} {76}},\ \bibinfo {pages} {033805} (\bibinfo
  {year} {2007}{\natexlab{c}})}\BibitemShut {NoStop}%
\bibitem [{\citenamefont {Scully}\ and\ \citenamefont {Zubairy}(1999)}]{HLM1}%
  \BibitemOpen
  \bibfield  {author} {\bibinfo {author} {\bibfnamefont {M.~O.}\ \bibnamefont
  {Scully}}\ and\ \bibinfo {author} {\bibfnamefont {M.~S.}\ \bibnamefont
  {Zubairy}},\ }\href@noop {} {\enquote {\bibinfo {title} {Quantum optics},}\ }
  (\bibinfo {year} {1999})\BibitemShut {NoStop}%
\bibitem [{\citenamefont {Gardiner}\ and\ \citenamefont {Zoller}(2004)}]{HLM2}%
  \BibitemOpen
  \bibfield  {author} {\bibinfo {author} {\bibfnamefont {C.}~\bibnamefont
  {Gardiner}}\ and\ \bibinfo {author} {\bibfnamefont {P.}~\bibnamefont
  {Zoller}},\ }\href@noop {} {\emph {\bibinfo {title} {Quantum noise: a
  handbook of Markovian and non-Markovian quantum stochastic methods with
  applications to quantum optics}}}\ (\bibinfo  {publisher} {Springer Science
  \& Business Media},\ \bibinfo {year} {2004})\BibitemShut {NoStop}%
\bibitem [{\citenamefont {Shaham}\ \emph {et~al.}(2020)\citenamefont {Shaham},
  \citenamefont {Katz},\ and\ \citenamefont {Firstenberg}}]{shaham2020quantum}%
  \BibitemOpen
  \bibfield  {author} {\bibinfo {author} {\bibfnamefont {R.}~\bibnamefont
  {Shaham}}, \bibinfo {author} {\bibfnamefont {O.}~\bibnamefont {Katz}}, \ and\
  \bibinfo {author} {\bibfnamefont {O.}~\bibnamefont {Firstenberg}},\
  }\href@noop {} {\bibfield  {journal} {\bibinfo  {journal} {Physical Review
  A}\ }\textbf {\bibinfo {volume} {102}},\ \bibinfo {pages} {012822} (\bibinfo
  {year} {2020})}\BibitemShut {NoStop}%
\bibitem [{\citenamefont {Lefevre-Seguin}\ and\ \citenamefont
  {Leduc}(1977)}]{lefevre1977metastability}%
  \BibitemOpen
  \bibfield  {author} {\bibinfo {author} {\bibfnamefont {V.}~\bibnamefont
  {Lefevre-Seguin}}\ and\ \bibinfo {author} {\bibfnamefont {M.}~\bibnamefont
  {Leduc}},\ }\href@noop {} {\bibfield  {journal} {\bibinfo  {journal} {Journal
  of Physics B: Atomic and Molecular Physics}\ }\textbf {\bibinfo {volume}
  {10}},\ \bibinfo {pages} {2157} (\bibinfo {year} {1977})}\BibitemShut
  {NoStop}%
\bibitem [{\citenamefont {Xia}\ \emph {et~al.}(2010)\citenamefont {Xia},
  \citenamefont {Morgan}, \citenamefont {Jau},\ and\ \citenamefont
  {Happer}}]{xia2010polarization}%
  \BibitemOpen
  \bibfield  {author} {\bibinfo {author} {\bibfnamefont {T.}~\bibnamefont
  {Xia}}, \bibinfo {author} {\bibfnamefont {S.}~\bibnamefont {Morgan}},
  \bibinfo {author} {\bibfnamefont {Y.-Y.}\ \bibnamefont {Jau}}, \ and\
  \bibinfo {author} {\bibfnamefont {W.}~\bibnamefont {Happer}},\ }\href@noop {}
  {\bibfield  {journal} {\bibinfo  {journal} {Physical Review A}\ }\textbf
  {\bibinfo {volume} {81}},\ \bibinfo {pages} {033419} (\bibinfo {year}
  {2010})}\BibitemShut {NoStop}%
\bibitem [{\citenamefont {Gentile}\ \emph
  {et~al.}(2017{\natexlab{b}})\citenamefont {Gentile}, \citenamefont {Nacher},
  \citenamefont {Saam},\ and\ \citenamefont {Walker}}]{gentile2017optically}%
  \BibitemOpen
  \bibfield  {author} {\bibinfo {author} {\bibfnamefont {T.~R.}\ \bibnamefont
  {Gentile}}, \bibinfo {author} {\bibfnamefont {P.}~\bibnamefont {Nacher}},
  \bibinfo {author} {\bibfnamefont {B.}~\bibnamefont {Saam}}, \ and\ \bibinfo
  {author} {\bibfnamefont {T.}~\bibnamefont {Walker}},\ }\href@noop {}
  {\bibfield  {journal} {\bibinfo  {journal} {Reviews of Modern Physics}\
  }\textbf {\bibinfo {volume} {89}},\ \bibinfo {pages} {045004} (\bibinfo
  {year} {2017}{\natexlab{b}})}\BibitemShut {NoStop}%
\bibitem [{\citenamefont {Batz}\ \emph {et~al.}(2011)\citenamefont {Batz},
  \citenamefont {Nacher},\ and\ \citenamefont
  {Tastevin}}]{batz2011fundamentals}%
  \BibitemOpen
  \bibfield  {author} {\bibinfo {author} {\bibfnamefont {M.}~\bibnamefont
  {Batz}}, \bibinfo {author} {\bibfnamefont {P.-J.}\ \bibnamefont {Nacher}}, \
  and\ \bibinfo {author} {\bibfnamefont {G.}~\bibnamefont {Tastevin}},\ }in\
  \href@noop {} {\emph {\bibinfo {booktitle} {Journal of Physics: Conference
  Series}}},\ Vol.\ \bibinfo {volume} {294}\ (\bibinfo {organization} {IOP
  Publishing},\ \bibinfo {year} {2011})\ p.\ \bibinfo {pages}
  {012002}\BibitemShut {NoStop}%
\bibitem [{\citenamefont {Shaham}\ \emph {et~al.}(2022)\citenamefont {Shaham},
  \citenamefont {Katz},\ and\ \citenamefont
  {Firstenberg}}]{Firstenberg-strong-coupling}%
  \BibitemOpen
  \bibfield  {author} {\bibinfo {author} {\bibfnamefont {R.}~\bibnamefont
  {Shaham}}, \bibinfo {author} {\bibfnamefont {O.}~\bibnamefont {Katz}}, \ and\
  \bibinfo {author} {\bibfnamefont {O.}~\bibnamefont {Firstenberg}},\
  }\href@noop {} {\bibfield  {journal} {\bibinfo  {journal} {Nature Physics}\
  }\textbf {\bibinfo {volume} {18}},\ \bibinfo {pages} {506} (\bibinfo {year}
  {2022})}\BibitemShut {NoStop}%
\bibitem [{\citenamefont {Gorshkov}\ \emph {et~al.}(2008)\citenamefont
  {Gorshkov}, \citenamefont {Calarco}, \citenamefont {Lukin},\ and\
  \citenamefont {S{\o}rensen}}]{Gorshkov4}%
  \BibitemOpen
  \bibfield  {author} {\bibinfo {author} {\bibfnamefont {A.~V.}\ \bibnamefont
  {Gorshkov}}, \bibinfo {author} {\bibfnamefont {T.}~\bibnamefont {Calarco}},
  \bibinfo {author} {\bibfnamefont {M.~D.}\ \bibnamefont {Lukin}}, \ and\
  \bibinfo {author} {\bibfnamefont {A.~S.}\ \bibnamefont {S{\o}rensen}},\
  }\href@noop {} {\bibfield  {journal} {\bibinfo  {journal} {Physical Review
  A}\ }\textbf {\bibinfo {volume} {77}},\ \bibinfo {pages} {043806} (\bibinfo
  {year} {2008})}\BibitemShut {NoStop}%
\bibitem [{\citenamefont {Heil}\ \emph {et~al.}(2013)\citenamefont {Heil},
  \citenamefont {Gemmel}, \citenamefont {Karpuk}, \citenamefont {Sobolev},
  \citenamefont {Tullney}, \citenamefont {Allmendinger}, \citenamefont
  {Schmidt}, \citenamefont {Burghoff}, \citenamefont {Kilian}, \citenamefont
  {Knappe-Gr{\"u}neberg} \emph {et~al.}}]{Heil-Noble-T2}%
  \BibitemOpen
  \bibfield  {author} {\bibinfo {author} {\bibfnamefont {W.}~\bibnamefont
  {Heil}}, \bibinfo {author} {\bibfnamefont {C.}~\bibnamefont {Gemmel}},
  \bibinfo {author} {\bibfnamefont {S.}~\bibnamefont {Karpuk}}, \bibinfo
  {author} {\bibfnamefont {Y.}~\bibnamefont {Sobolev}}, \bibinfo {author}
  {\bibfnamefont {K.}~\bibnamefont {Tullney}}, \bibinfo {author} {\bibfnamefont
  {F.}~\bibnamefont {Allmendinger}}, \bibinfo {author} {\bibfnamefont
  {U.}~\bibnamefont {Schmidt}}, \bibinfo {author} {\bibfnamefont
  {M.}~\bibnamefont {Burghoff}}, \bibinfo {author} {\bibfnamefont
  {W.}~\bibnamefont {Kilian}}, \bibinfo {author} {\bibfnamefont
  {S.}~\bibnamefont {Knappe-Gr{\"u}neberg}},  \emph {et~al.},\ }\href@noop {}
  {\bibfield  {journal} {\bibinfo  {journal} {Annalen der Physik}\ }\textbf
  {\bibinfo {volume} {525}},\ \bibinfo {pages} {539} (\bibinfo {year}
  {2013})}\BibitemShut {NoStop}%
\bibitem [{\citenamefont {Budker}\ and\ \citenamefont
  {Romalis}(2007)}]{Budker-Romalis-Magnetometry}%
  \BibitemOpen
  \bibfield  {author} {\bibinfo {author} {\bibfnamefont {D.}~\bibnamefont
  {Budker}}\ and\ \bibinfo {author} {\bibfnamefont {M.}~\bibnamefont
  {Romalis}},\ }\href@noop {} {\bibfield  {journal} {\bibinfo  {journal}
  {Nature Physics}\ }\textbf {\bibinfo {volume} {3}},\ \bibinfo {pages} {227}
  (\bibinfo {year} {2007})}\BibitemShut {NoStop}%
\bibitem [{\citenamefont {Rumelhart}\ \emph {et~al.}(1986)\citenamefont
  {Rumelhart}, \citenamefont {Hinton},\ and\ \citenamefont
  {Williams}}]{gradient-ascent-1}%
  \BibitemOpen
  \bibfield  {author} {\bibinfo {author} {\bibfnamefont {D.~E.}\ \bibnamefont
  {Rumelhart}}, \bibinfo {author} {\bibfnamefont {G.~E.}\ \bibnamefont
  {Hinton}}, \ and\ \bibinfo {author} {\bibfnamefont {R.~J.}\ \bibnamefont
  {Williams}},\ }\href@noop {} {\bibfield  {journal} {\bibinfo  {journal}
  {Nature}\ }\textbf {\bibinfo {volume} {323}},\ \bibinfo {pages} {533}
  (\bibinfo {year} {1986})}\BibitemShut {NoStop}%
\bibitem [{\citenamefont {Qian}(1999)}]{gradient-ascent-2}%
  \BibitemOpen
  \bibfield  {author} {\bibinfo {author} {\bibfnamefont {N.}~\bibnamefont
  {Qian}},\ }\href@noop {} {\bibfield  {journal} {\bibinfo  {journal} {Neural
  Networks}\ }\textbf {\bibinfo {volume} {12}},\ \bibinfo {pages} {145}
  (\bibinfo {year} {1999})}\BibitemShut {NoStop}%
\end{thebibliography}%

\end{document}